\documentclass[12pt]{article}

\usepackage{amsmath}
\usepackage{feynmf}
\usepackage{latexsym}
\usepackage{amssymb}
\usepackage[dvips]{graphicx}
\usepackage{caption}
\usepackage{rotating}
\usepackage{latexsym}
\usepackage{verbatim}

 \def\be{\begin{equation}}
 \def\ee{\end{equation}}
 \def\bea{\begin{eqnarray}}
 \def\eea{\end{eqnarray}}
 \def\bei{\begin{itemize}}
 \def\eei{\end{itemize}}
 \def\bs{\begin{slide}}
 \def\es{\end{slide}}
 \def\nn{\nonumber}
 \def\half{\frac{1}{2}}
 \def\pd{\partial}
 
 \def\L{\mathcal{L}}
 
 \def\a{\alpha}
 \def\ad{{\dot\alpha}}
 \def\b{\beta}
 \def\bd{{\dot\beta}}
 \def\g{\gamma}
 \def\G{\Gamma}
 \def\d{\delta}
 \def\D{\Delta}
 \def\m{\mu}
 \def\n{\nu}
 \def\t{\tau}
 \def\r{\rho}
 \def\th{\theta}
 \def\w{\omega}
 
 \def\l{\lambda}
 \def\lb{{\bar\l}}
 
 \def\s{\sigma}
 
 \def\sb{\bar{\s}}
 \def\psib{{\bar{\psi}}}
 \def\e{\epsilon}
 \def\f{\phi}

 \def\MZp{M_{Z'}}
 \def\MZO{M_{Z_0}}
 
 \def\({\left(}
 \def\){\right)}
 \def\[{\left[}
 \def\]{\right]}
 
 \def\gt{\tilde{g}}
 \def\Qt{\tilde{Q}}
 \def\Ut{{{\tilde{U}^c}}}
 \def\Dt{{{\tilde{D}^c}}}
 \def\Uts{{{\tilde{U}^{c\dag}}}}
 \def\Dts{{{\tilde{D}^{c\dag}}}}
 \def\Lt{\tilde{L}}
 \def\Et{{{\tilde{E}^c}}}
 \def\Ets{{{\tilde{E}^{c\dag}}}}

 \def\la{\langle}
 \def\ra{\rangle}
 \def\Tr{\textnormal{Tr}}

 \def\Eps{\epsilon^{\mu\nu\rho\sigma}}
 \def\emn{\eta_{\m\n}}

 \def\thth{\theta^2 \bar\theta^2}
 
 \def\QQ{{Q_Q}}
 \def\QU{{Q_{U^c}}}
 \def\QD{{Q_{D^c}}}
 \def\QL{{Q_L}}
 \def\QE{{Q_{E^c}}}
 \def\QHu{{Q_{H_u}}}
 \def\QHd{{Q_{H_d}}}
 \def\cA{\mathcal{A}}

 \newcommand{\bth}{{\bf 3}}
 \newcommand{\btw}{{\bf 2}}
 \newcommand{\bon}{{\bf 1}}
 \def\slashed{\ds}
 \def\ds#1{#1\kern-1ex\hbox{/}}
\def\sla{\raise.15ex\hbox{$/$}\kern-.57em}

\begin{document}

\begin{titlepage}

\rightline{0804.1156 [hep-th]}

\rightline{ROM2F/2008/11}

\vskip 2cm

\centerline{{\large\bf Minimal Anomalous U(1)' Extension of the MSSM}}

\vskip 1cm

\centerline{Pascal Anastasopoulos\footnote{Pascal.Anastasopoulos@roma2.infn.it},~
Francesco Fucito\footnote{Francesco.Fucito@roma2.infn.it},~
Andrea Lionetto\footnote{Andrea.Lionetto@roma2.infn.it},}
\vskip .2 cm
\centerline{Gianfranco Pradisi\footnote{Gianfranco.Pradisi@roma2.infn.it},~
Antonio Racioppi\footnote{Antonio.Racioppi@roma2.infn.it},~
Yassen S. Stanev\footnote{Yassen.Stanev@roma2.infn.it}}

\vskip 1cm

\centerline{Dipartimento di Fisica dell'Universit\`a di Roma , ``Tor Vergata" and}
\centerline{I.N.F.N.~ -~ Sezione di Roma ~ ``Tor Vergata''}
\centerline{Via della Ricerca  Scientifica, 1 - 00133 ~ Roma,~ ITALY}

\begin{abstract}

We study an extension of the MSSM by an anomalous abelian vector multiplet and a St\"uckelberg multiplet.
The anomalies are cancelled by the Green-Schwarz mechanism and the addition of Chern-Simons terms.
The advantage of this choice over the standard one is that it allows for arbitrary values of the quantum numbers
of the extra $U(1)$. As a first step towards the study of hadron annihilations producing four leptons in the final
state (a clean signal which might be studied at LHC) we then compute the decays $Z'\to Z_0 \g$ and $Z'\to Z_0 Z_0$.
We find that the largest values of the decay rate is $\sim 10^{-4}$ GeV, while the expected number of
events per year at LHC is at most of the order of 10.

\end{abstract}

\end{titlepage}

\tableofcontents

\section{Introduction}

The Standard Model (SM) of particle physics has been confirmed to
a great accuracy in many experiments. Despite the fact that the
Higgs particle remains experimentally elusive, few scientists
doubt that there will be major surprises in this direction. The
whole scientific community, however, knows that the SM needs to be
improved. First of all, neutrino oscillation experiments have
exhibited the evidence for (tiny) neutrino masses, that have to be
incorporated in (an extension of) the SM. Many ideas exist on how
this can be achieved and more experimental precision tests will
indicate which models are viable. Second, there are also several
theoretical issues that make physicists believe that the SM is
only an effective manifestation of a more Fundamental Theory.

In approximately one year, the Large Hadron Collider (LHC) at CERN
will start to operate at energies of order of 14 TeV in the center of
mass. Apart from the search for the Higgs boson,
it will probably give us some answers about the parameter space of the
physics beyond the SM. Among the many issues that will be
addressed, it is worth to mention: the search for supersymmetry,
heavy quarks and the quark-gluon plasma, the existence of extra
dimensions and the possible creation of tiny black holes.

One of the most attractive scenario for physics beyond the SM is
the existence of additional massive neutral gauge bosons
\cite{Leike:1998wr}-\cite{Saxena:2007xx}.  They could be one of the first
discoveries at LHC if their
mass is in the range of a few TeV.
Many different models
have been developed in the past in order to investigate this
possibility. The mass could be acquired
in a variety of ways: from Kaluza-Klein modes to a standard
Higgs mechanism or even by adding an axionic field, $\f$, which
couples to the abelian factors (St\"uckelberg mechanism)
\cite{Kors:2004dx,Feldman:2007wj}. The latter is common
to low energy effective field theories which
appear anomalous.
The anomaly cancellation is
achieved by the Green-Schwarz mechanism with St\"uckelberg terms
accompanied by axion like couplings, $\f F\tilde{F}$, which
ensure the
consistency of these models \cite{Ibanez:1999it, Coriano':2005js}.

For example, in string theory anomalous $U(1)$'s are very common.
D-brane models contain several abelian factors, living on each
stack of branes, and they are typically anomalous
\cite{Pradisi:1988xd}-\cite{Kiritsis:2007zz}. In the presence of
these anomalous $U(1)$'s, the St\"uckelberg mixing with the axions
cancels mixed anomalies\footnote{Irreducible anomalies are
cancelled by the tadpole cancellation.} \cite{Ibanez:1998qp}, and renders the ``anomalous" gauge fields massive.
The masses depend non-trivially on the internal volumes and on
other moduli, allowing the physical masses of the anomalous $U(1)$
gauge bosons to be much smaller than the string scale (even at a
few TeV range) \cite{Ibanez:1999it, Antoniadis:2002cs}.
However, it has been shown that axionic terms alone are not
sufficient to cancel all anomalies. An important role is played by
the so-called Generalized Chern-Simons terms (GCS) which are local
gauge non-invariant terms. Indeed, these trilinear gauge bosons anomalous couplings
are responsible for the cancellation of mixed anomalies between
anomalous $U(1)'$s and non anomalous factors ensuring the
consistency of the theory \cite{de Wit:1984px}-\cite{DeRydt:2007vg}.

In this paper, we are interested in {\it anomaly related} $Z'$ bosons in a non-renormalizable effective field theory. More precisely, we study an extension of
the MSSM (see \cite{Martin:1997ns} for a review) by the addition of an abelian vector multiplet $V^{(0)}$ and we assume that generically all MSSM
particles are charged with respect to the new $U(1)$. In order to gain in flexibility, our model is only string
inspired: we do not commit to a specific brane model and this is why the charges are not fixed, even if the effective cut-off is related to the mass of the $Z'$.  The extra vector
multiplet generically is anomalous and consistency of the model requires an additional St\"uckelberg multiplet $S$
with the proper couplings as well as GCS terms. As a consequence, the anomalous abelian boson becomes massive and
behaves like a $Z'$.
Moreover, in order to break supersymmetry, we add the usual soft breaking terms and the new terms coming from the
fermionic sectors of $V^{(0)}$ and $S$.

Our model contains many new features: new D and F terms (which
are coming from the axionic terms and not from the GCS, in
accordance with \cite{DeRydt:2007vg}, due to the fact that the
GCS's contain only vector multiplets in antisymmetric form), new
couplings and new mass contributions in comparison with the
MSSM. Explicit formulae are provide for all these
terms in component fields.

Since the Higgs fields might be charged under the anomalous
$U(1)$, a combination of the St\"uckelberg and the Higgs mechanism
makes the anomalous $U(1)$ massive. An axi-Goldstone combination
is eaten by the neutral gauge bosons and no
physical axi-Higgs is left contrary to other studies on anomaly related
$Z'$ \cite{Coriano':2005js} and similarly to the case of a
non-anomalous related $Z'$ \cite{Kors:2004dx,Feldman:2007wj}.

We explicitly show how the anomaly cancellation mechanism works in
our model before and after breaking the gauge symmetry. Before
gauge symmetry breaking, only SM fermions contribute to the
triangle diagrams. After gauge symmetry breaking, all fermions
that become massive still contribute to the anomalous triangle
diagrams. Their contribution is cancelled by new diagrams which
involve the Nambu-Goldstone (NG) boson exchange.

In order to explore some phenomenological implications of our
setting, we then analyze the decays $Z'\to Z_0\gamma$ and $Z'\to
Z_0Z_0$. We numerically compute the decay rates as functions of the arbitrary
$U(1)$ charges and the mass of the anomalous $U(1)$ gauge boson.
We find a non-trivial dependence on all these parameters,
estimating that the region that gives the largest values is for $M_{Z'}\sim 4$ TeV,
where the decay rate $Z'\to Z_0\gamma$ is of the order of $10^{-4}$ GeV.
These decays are part of the processes in which two colliding
protons lead to a four lepton final state \cite{CDF}. The final state is very
clean and possibly measurable at LHC.
Assuming a degenerate mass spectrum for the sfermions of about 500 GeV we also estimate $N_{Z'}$, the
expected number of $Z'$ produced per year. We find that $N_{Z'}$ falls off
exponentially with $M_{Z'}$, so we shall focus on the case $M_{Z'} \sim 1$
TeV and the most favorite decay $Z'\to Z_0 Z_0$. We also estimate the number of decays for 1 year
of integrated luminosity which turns out to be $N_{Z'\to Z_0 Z_0} \sim 10$
in the most favourite region of parameters.
In a future
work we will push our program forward and study this signal with the
aid of Monte Carlo methods \cite{inprogress}.

The paper is organized as follows: in Section \ref{setup}, we introduce the vector multiplet, $V^{(0)}$,
the St\"uckelberg multiplet and we
provide the axionic and GCS lagrangians in superfields and in components.
We then discuss the anomaly cancellation
both in the unbroken and in the broken phase. At the end of the
Section, we add all possible soft-breaking terms.
In Section \ref{Consequences}, we describe the model set up. In particular, we discuss the
kinetic mixing terms which are coming from the axionic lagrangian and the D and F  terms, pointing
out explicitly the new contributions. We comment on the superpotential and we compute the mass terms for all the
particles, pointing out the differences from the canonical MSSM setup.
Finally, in Section \ref{decays}, we study some phenomenomogical
implications of our model. We consider the case in which the Higgs fields are uncharged with respect to the $U(1)'$ and compute the decay rates
for the two processes $Z'\to
Z_0\gamma$ and $Z'\to Z_0Z_0$ which should be relevant for the computation of hadron annihilations into four leptons.
In the appendices we report the technical
details and discuss the general case in which also the Higgs
fields transform under the anomalous $U(1)'$.

\section{Preliminaries}\label{setup}

In this section, we discuss how to extend the Minimal Supersymmetric Standard Model (MSSM) to accommodate an
additional abelian vector multiplet $V^{(0)}$ and how to cancel the anomalies with the Green-Schwarz mechanism. We
assume that all the MSSM fields are charged under the additional vector multiplet $V^{(0)}$, with charges that are
given in Table~\ref{QTable}, where $Q_i, L_i$ are the left handed quarks and leptons respectively while $U^c_i,
D^c_i, E^c_i$ are the right handed up and down quarks and the electrically charged leptons. The superscript $c$ stands
for charge conjugation. The index $i=1,2,3$ denotes the three different families. $H_{u,d}$ are the two Higgs
scalars.
  \begin{table}[h]
  \centering
  \begin{tabular}[h]{|c|c|c|c|c|}
   \hline & SU(3)$_c$ & SU(2)$_L$  & U(1)$_Y$ & ~U(1)$^{\prime}~$\\
   \hline $Q_i$   & $\bth$       &  $\btw$       &  $1/6$   & $Q_{Q}$ \\
   \hline $U^c_i$   & $\bar \bth$  &  $\bon$       &  $-2/3$  & $Q_{U^c}$ \\
   \hline $D^c_i$   & $\bar \bth$  &  $\bon$       &  $1/3$   & $Q_{D^c}$ \\
   \hline $L_i$   & $\bon$       &  $\btw$       &  $-1/2$  & $Q_{L}$ \\
   \hline $E^c_i$   & $\bon$       &  $\bon$       &  $1$     & $Q_{E^c}$\\
   \hline $H_u$ & $\bon$       &  $\btw$       &  $1/2$   & $Q_{H_u}$\\
   \hline $H_d$ & $\bon$       &  $\btw$       &  $-1/2$  & $Q_{H_d}$ \\
   \hline
  \end{tabular}
  \caption{Charge assignment.}\label{QTable}
  \end{table}

Since our model is an extension of the MSSM, the gauge invariance
of the superpotential, that contains the Yukawa couplings and a
$\m$-term, put constraints on the above charges
  \bea
   \QU &=& - \QQ - \QHu  \nn\\
   \QD &=& - \QQ + \QHu  \nn\\
   \QE &=& -\QL  + \QHu  \nn\\
   \QHd  &=& - \QHu \label{Qconstraints}
  \eea
Thus, $\QQ$, $\QL$ and $\QHu$ are free parameters of the model.

\subsection{Anomalies} \label{anomalies}

As it is well known, the MSSM is anomaly free. All the anomalies
that involve only the $SU(3)$, $SU(2)$ and $U(1)_Y$ factors vanish
identically. However, triangles with $U(1)'$ in the external legs
in general are potentially anomalous. These anomalies are\footnote{We are working in an effective field theory framework and we ignore troughout the paper all the gravitational effects.  In particular, we do not consider the gravitational anomalies which, however, could be canceled by the Green-Schwarz mechanism.}
\bea
   U(1)'-U(1)'-U(1)'~~~:     &&\ \cA^{(0)} = \sum_f Q_f^3                        \label{Triangles1}\\
   U(1)'-U(1)_Y - U(1)_Y~~~: &&\ \cA^{(1)} = \sum_f Q_f Y_f^2                     \label{Triangles2}\\
   U(1)'-SU(2)-SU(2)~~~:     &&\ \cA^{(2)} = \sum_f Q_f \Tr[T_{k_2}^{(2)} T_{k_2}^{(2)}] \label{Triangles3}\\
   U(1)'-SU(3)-SU(3)~~~:     &&\ \cA^{(3)} = \sum_f Q_f \Tr[T_{k_3}^{(3)} T_{k_3}^{(3)}] \label{Triangles4}\\
   U(1)'-U(1)'-U(1)_Y~~~:    &&\ \cA^{(4)} = \sum_f Q_f^2  Y_f
   \label{Triangles5}
\eea
where $f$ runs over the fermions in Table \ref{QTable}, $Q_f$ is
the corresponding $U(1)'$ charge, $Y_f$ is the hypercharge and
$T_{k_a}^{(a)}$, $a=2,3;\,\, k_a=1,\ldots,{\rm dim G}^{(a)}$ are
the generators of the $G^{(2)}=SU(2)$ and $G^{(3)}=SU(3)$ algebras
respectively. In our notation $\Tr[T_j^{(a)} T_k^{(a)}] = {1\over
2}\d_{jk}$. All the remaining anomalies that involve $U(1)'$s
vanish identically due to group theoretical arguments
(see Chapter 22 of \cite{Weinberg2}). Using the charge constraints
(\ref{Qconstraints}) we get
  \bea
   \cA^{(0)} &=& 3\ \Big\{ Q_{H_u}^3 + 3 \QHu Q_L^2 + Q_L^3 - 3 Q_{H_u}^2\ \( \QL + 6 \QQ \) \Big\} \label{A0}\\
   \cA^{(1)} &=& -{3\over2} \(3\QQ + \QL  \) \label{A1}\\
   \cA^{(2)} &=&  {3\over2} \(3\QQ  + \QL \) \label{A2}\\
   \cA^{(3)} &=& 0 \label{A3}\\
   \cA^{(4)} &=& -6 \QHu \(3\QQ + \QL  \)
 \label{A4} \eea
Notice that the mixed anomaly between the anomalous $U(1)$ and the $SU(3)$ nonabelian factors $\cA^{(3)}$ vanishes
identically.

\subsubsection{Anomalous U(1)'s and the St\"uckelberg mechanism}

Many models have been developed in the past where all the anomalies (\ref{A0}-\ref{A4}) vanish by constraining the
charges $Q_f$ (see \cite{Leike:1998wr, Yao:2006px} and references therein). On the contrary, in this paper we
assume that the $U(1)'$ is anomalous, i.e. (\ref{A0})-(\ref{A4}) do not vanish. Consistency of the model is achieved
by the contribution of a St\"uckelberg field $S$ and its appropriate couplings to the anomalous $U(1)'$. The
St\"uckelberg lagrangian reads \cite{Klein:1999im}
  \bea
   \L_{axion} &=& {1\over4} \left. \( S + S^\dagger + 4 b_3 V^{(0)} \)^2 \right|_{\thth} \nn\\
                &&- {1\over2} \left\{ \[\sum_{a=0}^2 b^{(a)}_2 S ~\Tr\( W^{(a)} W^{(a)} \) + b^{(4)}_2 S ~W^{(1)} ~W^{(0)} \]_{\th^2} +h.c. \right\}~
   \label{Laxion}\eea
where the index $a=0,\ldots,3$ runs over the $U(1)',\, U(1)_Y,\, SU(2)$ and $SU(3)$ gauge groups respectively.
The St\"uckelberg multiplet is a chiral superfield
 \be
    S =  s+ i\sqrt2 \th \psi_S + \th^2 F_S - i \th \s^\m \bar\th \pd_\m s +
                {\sqrt2\over2}  \th^2 \bar\th \bar\s^\m \pd_\m \psi_S - {1\over4} \thth \Box s \label{Smult}
 \ee
and transforms under the $U(1)'$ as
\bea
   V^{(0)} &\to& V^{(0)} + i \( \Lambda - \Lambda^\dag \) \nn\\
   S  &\to& S - 4 i ~b_3 ~\Lambda \label{U1'}
  \label{U1Trans}\eea
where $b_3$ is a constant. The lowest component
of $S$ is a complex scalar field $s=\a+i \f $. We assume that
the real part $\a$ gets an expectation value by an effective
potential of stringy or different origin and contributes to the
coupling constants as
  \be
    \frac{1}{16 g_a^2 \t_a}=\frac{1}{16 \gt_a^2 \t_a} -{1\over2} b^{(a)}_2 \langle \a\rangle \label{couplingconst}
   \ee
where $g_a$ is the redefined coupling constant and the gauge factors $\t_a$ take the values $1,1,1/2,1/2$.
The first line in (\ref{Laxion}) is gauge invariant and provides
the kinetic terms and the axion-$U(1)'$ mixing. The second line is
not gauge invariant and provides couplings that participate in the
anomaly cancellation procedure. Notice that in (\ref{Laxion}) the
sum over $a$ omits the $a=3$ case since there is no mixed
anomaly between the $U(1)'$ and the $SU(3)$ factors as from eq.(\ref{A3}),
i.e. $b_2^{(3)}=0$. The values of the other constants,
$b_2^{(a)}$, are fixed by the anomalies.

At first sight our lagrangian (see Appendix~\ref{applagrangian}) may look not the most general possible one.
In particular, an explicit Fayet-Iliopoulos term $\xi V^{(0)}$
could be added.
It is well known that in certain string-inspired models (see, e.g.
\cite{Poppitz:1998dj}), an one-loop FI term is absent, even if
$Tr(Q) \neq 0$. This  is in  apparent conflict with the observation
\cite{Fischler:1981zk} that in field theory a quadratically divergent FI term is
always generated at one loop.
The solution to this paradox is that in the low-energy lagrangian there
should be a counterterm, which compensates precisely, i.e. both the
divergent and the finite part of, the one loop contribution.
We do not write explicitly this counterterm, since its exact expression is
model and regularization dependent, but we implicitly assume that such a
cancellation occurs. As mentioned before, also the terms responsible for the cancellation of gravitational anomalies are omitted.

Expanding $\L_{axion}$ in component fields, using the Wess-Zumino gauge and substituting $\a$ by its vev we get
   \bea
    \L_{axion} &=& {1\over2} \( \pd_\m \f +2 b_3 V^{(0)}_\m \)^2
                   +{i\over4} \psi_S \s^\m \pd_\m \psib_S +{i\over4} \psib_S \sb^\m \pd_\m \psi_S \\
                 && +{1\over2} F_S \bar F_S + 2 b_3 \langle\a\rangle D^{(0)}-\sqrt2 b_3(\psi_S \l^{(0)}+h.c.)\nn\\
                 &&- {1\over4} \f \, \Eps \sum_{a=0}^2 b^{(a)}_2 \Tr \(  F_{\m \n}^{(a)} F_{\r \s}^{(a)} \)- {1\over4} b^{(4)}_2 \Eps \f F_{\m \n}^{(1)} F_{\r \s}^{(0)}\nn\\
                 &&+{1\over2} b^{(4)}_2 \langle\a\rangle F_{\m \n}^{(1)} F_{\m \n}^{(0)}- b^{(4)}_2  \langle\a\rangle D^{(1)} D^{(0)} \nn\\
                 &&-{1\over2} \left\{\sum_{a=0}^2b^{(a)}_2 \[- 2\f \Tr \( \l^{(a)} \s^\m D_\m \lb^{(a)} \) +
                       {i\over\sqrt2} \Tr \( \l^{(a)} \s^\m \sb^\n F_{\m \n}^{(a)} \) \psi_S \right. \right. \nn\\
                 &&\left. - F_S \Tr \(\l^{(a)} \l^{(a)}\) - \sqrt2 \psi_S \Tr \(\l^{(a)} D^{(a)}\)\]\nn\\
                 &&+ b^{(4)}_2 \bigg[ \(-\f \l^{(1)} \s^\m \pd_\m \lb^{(0)}
                 +i\langle\a\rangle  \l^{(1)} \s^\m \pd_\m \lb^{(0)}  -{1\over2}F_S \l^{(1)} \l^{(0)}\right.\nn\\
                 && \left. \left.- {1\over\sqrt2} \psi_S  \l^{(1)} D^{(0)}
                 +{i\over2\sqrt2} \l^{(1)} \s^\m \sb^\n F_{\m \n}^{(0)} \psi_S\)
                 + (0 \leftrightarrow 1) \ \bigg] +h.c. \right\}\nn
   \eea
where we omit terms which are coming from $\la \a \ra W^{(a)}
W^{(a)}$, since they are absorbed in the coupling constant
redefinition (\ref{couplingconst}). This mechanism cancels some
mixed anomalies and in addition provides a mass term to the
anomalous $U(1)$. Therefore, the anomalous $U(1)$ behaves $almost$
like the usual $Z'$ extensively studied in the past.

\subsubsection{Generalized Chern-Simons terms}

As it was pointed out in \cite{Anastasopoulos:2006cz}, the
St\"uckelberg mechanism is not sufficient to cancel all the
anomalies. Mixed anomalies between anomalous and non-anomalous
factors require an additional mechanism to ensure consistency of
the model: non gauge invariant Generalized Chern-Simons terms
(GCS) must be added.
In our case, the GCS terms have the form
\cite{Andrianopoli:2004sv}
   \bea
    \L_{GCS} &=&- d_4     \[ \( V^{(1)} D^\a V^{(0)} - V^{(0)} D^\a V^{(1)}\) W^{(0)}_\a + h.c. \]_{\thth} +\nn\\
             &&+  d_5     \[ \( V^{(1)} D^\a V^{(0)} - V^{(0)} D^\a V^{(1)}\) W^{(1)}_\a + h.c. \]_{\thth} +\nn\\
             &&+  d_6 \Tr \bigg[ \( V^{(2)} D^\a V^{(0)} - V^{(0)} D^\a V^{(2)}\) W^{(2)}_\a +\nn\\
                          &&\qquad \quad+{1\over6} V^{(2)} D^\a V^{(0)} \bar D^2 \(  \[D_\a V^{(2)},V^{(2)}\] \) + h.c. \bigg]_{\thth}
   \label{GCS_1} \eea
The constants $d_4$, $d_5$ and $d_6$ are fixed by the cancellation of the mixed anomalies. The GCS terms
(\ref{GCS_1}), expressed in component fields, are
   \bea
    \L_{GCS} &=& -d_4 ~\Eps V^{(0)}_\m V^{(1)}_\n F_{\r \s}^{(0)} + d_5 ~\Eps V^{(0)}_\m V^{(1)}_\n F_{\r \s}^{(1)}\nn\\
                &&+d_6 ~\Eps V^{(0)}_\m \, \Tr \[  V^{(2)}_\n F_{\r \s}^{(2)}
                  -{i\over3} V^{(2)}_\n \[V^{(2)}_\r, V^{(2)}_\s\] \] \nn\\
                &&-d_4 \( \l^{(0)} \s^\m \lb^{(0)} V^{(1)}_\m -  \l^{(0)} \s^\m \lb^{(1)} V^{(0)}_\m   + h.c.  \) \nn\\
                &&+d_5 \( \l^{(1)} \s^\m \lb^{(1)} V^{(0)}_\m  -  \l^{(1)} \s^\m \lb^{(0)} V^{(1)}_\m + h.c. \)\nn\\
                &&+d_6 \Tr \[\l^{(2)} \s^\m \lb^{(2)} V^{(0)}_\m - \l^{(2)} \s^\m \lb^{(0)} V^{(2)}_\m +h.c.\]
   \eea
These terms provide new trilinear
couplings that distinguish these models from the $Z'$ models
studied in the past.

\subsection{Anomaly cancellation}\label{AnomalyCancellation}

In the following, we illustrate the anomaly cancellation procedure both
in the unbroken and broken phases by a specific example. We focus
on the bosonic sector and the related diagrams, since their
supersymmetric analogs are fixed by supersymmetry.
The GS and GCS terms depend on unknown parameters which we fix by
using the Ward identities. In theories with massive gauge bosons
where the mass is acquired either by the Higgs or by the
Stuckelberg mechanism, Ward identities have the following
diagrammatic form \cite{Chanowitz:1985hj}
\vskip 0.25cm
\bea
-i k^\m~\Bigg(~~~~~~~~~
      \raisebox{-4.7ex}[0cm][0cm]{\unitlength=0.7mm
      \begin{fmffile}{vectorbubble5a}
      \begin{fmfgraph*}(40,25)
       \fmfpen{thick} \fmfleft{o1} \fmfright{i1,i2} \fmfstraight \fmftop{t1,t2,t3,t4,i2}\fmfbottom{b1,b2,b3,b4,i1}
       \fmf{boson}{o1,v1}   \fmf{boson}{v1,b4}       \fmf{boson}{v1,i2}
              \fmf{vanilla}{v1,t4}       \fmf{vanilla}{v1,i1}
              \fmffreeze     \fmf{boson}{v1,t3}
       \fmfv{decor.shape=circle,decor.filled=shaded, decor.size=.25w}{v1}
       \fmflabel{~~1PI}{v1} \fmflabel{$V^\m(k)$}{o1}
      \end{fmfgraph*}
      \end{fmffile}}~~~\Bigg)
  ~ + m_V~
     \Bigg(~~~~~~~~~
      \raisebox{-4.7ex}[0cm][0cm]{\unitlength=0.7mm
      \begin{fmffile}{goldbubble5a}
      \begin{fmfgraph*}(40,25)
       \fmfpen{thick} \fmfleft{o1} \fmfright{i1,i2} \fmfstraight \fmftop{t1,t2,t3,t4,i2}\fmfbottom{b1,b2,b3,b4,i1}
       \fmf{dashes}{o1,v1}    \fmf{boson}{v1,b4}       \fmf{boson}{v1,i2}
              \fmf{vanilla}{v1,t4}       \fmf{vanilla}{v1,i1}
              \fmffreeze    \fmf{boson}{v1,t3}
       \fmfv{decor.shape=circle,decor.filled=shaded,label=1PI , decor.size=.25w}{v1}
       \fmflabel{~~1PI}{v1} \fmflabel{$G_V(k)$}{o1}
      \end{fmfgraph*}
      \end{fmffile}}~~~\Bigg)
= 0 ~~~~~\label{GoldWI}\\ \nn
\eea
where $V_\m$ is the massive gauge field, $G_V$ is the
corresponding Higgs or St\"uckelberg field (or a linear
combination of them) and $m_V$ is the coupling of the term $V^\m
\pd_\m G_V$. The blob denotes all the 1PI diagrams.

\subsubsection{Anomaly cancellation in the symmetric phase}\label{AnomalyCancellationSymmPhase}

In our model there are two extra states in the neutral fermionic sector, namely the axino and the primeino (see Section \ref{Neutralinos}) which
do not contribute to the fermionic loop. The remaining MSSM fermionic states are a bino, a wino and the two higgsinos. Both $U(1)_Y$
and $SU(2)$ gauginos do not contribute to the fermionic loop due to group theoretical arguments (see Section 28.1 of \cite{Weinberg3}). The higgsino
eigenstates do not participate because the $\tilde{H}_u$ contribution is cancelled by the $\tilde{H}_d$ one. This is due to the fact that each
diagram is proportional to an odd product of charges and the two higgsinos have opposite charges (see Table \ref{QTable} and the constraints
(\ref{Qconstraints})).
Without loss of generality, we assume that the mixed anomaly between $V^{(0)}$ and two $V^{(1)}$ is non vanishing,
therefore from eq. (\ref{Triangles2}) $\cA^{(1)}=\sum_f Q_f (Y_f)^2\neq 0$. In order to cancel the anomaly, we have to
satisfy the Ward identities which are shown, in diagrammatic form, in Fig. \ref{DiagramsUnbroken}.
     \begin{figure}[tb]
  \vskip 1.5cm
%
$(p+q)^\r ~\Bigg($~~~~~~~~~~~~~~~
      \raisebox{-4ex}[0cm][0cm]{\unitlength=0.4mm
      \begin{fmffile}{pYYloop3}
      \begin{fmfgraph*}(60,40)
       \fmfpen{thick} \fmfleft{ii0,ii1,ii2} \fmfstraight \fmffreeze \fmftop{ii2,t1,t2,t3,oo2}
       \fmfbottom{ii0,b1,b2,b3,oo1} \fmf{phantom}{ii2,t1,t2} \fmf{phantom}{ii0,b1,b2} \fmf{phantom}{t1,v1,b1}
       \fmf{phantom}{t2,b2} \fmf{phantom}{t3,b3}
       \fmffreeze
       \fmf{photon}{ii1,v1} \fmf{fermion}{t3,v1} \fmf{fermion,label=$\psi$}{b3,t3} \fmf{fermion}{v1,b3}
       \fmf{photon}{b3,oo1} \fmf{photon}{t3,oo2}
       \fmflabel{$V^{(0)}_\r(p+q)$}{ii1}
       \fmflabel{$V^{(1)}_\m(p)$}{oo1}
       \fmflabel{$V^{(1)}_\n(q)$}{oo2}
      \end{fmfgraph*}
      \end{fmffile}}      ~~~~+~~~~~~~~
      \raisebox{-4ex}[0cm][0cm]{\unitlength=0.5mm
      \begin{fmffile}{GCS_pYY1bb}
      \begin{fmfgraph*}(30,30)
       \fmfpen{thick}
       \fmfleft{i1} \fmfright{o1,o2}
       \fmf{boson}{i1,v1} \fmf{boson}{v1,o1} \fmf{boson}{v1,o2}
       \fmffreeze
       \fmflabel{$V^{(0)}$}{i1}\fmflabel{$V^{(1)}$}{o1}\fmflabel{$V^{(1)}$}{o2}
      \end{fmfgraph*}
      \end{fmffile}} $\Bigg)$
$+2ib_3 \Bigg($
      \raisebox{-4ex}[0cm][0cm]{ \unitlength=0.5mm
      \begin{fmffile}{axionYY}
      \begin{fmfgraph*}(30,30)
      \fmfpen{thick} \fmfleft{i1} \fmfright{o1,o2}
      \fmf{dashes}{i1,v1} \fmf{photon}{v1,o2}
      \fmf{photon}{v1,o1} \fmffreeze \fmflabel{$V^{(1)}$}{o1} \fmflabel{$V^{(1)}$}{o2} \fmflabel{$\f$}{i1}
      \end{fmfgraph*}
      \end{fmffile}} $\Bigg)=0$
        \vskip 2cm

~~~~~~$p^\m ~\Bigg($~~~~
      \raisebox{-4ex}[0cm][0cm]{\unitlength=0.4mm
      \begin{fmffile}{pYYloop222}
      \begin{fmfgraph*}(60,40)
       \fmfpen{thick} \fmfleft{ii0,ii1,ii2} \fmfstraight \fmffreeze \fmftop{ii2,t1,t2,t3,oo2}
       \fmfbottom{ii0,b1,b2,b3,oo1} \fmf{phantom}{ii2,t1,t2} \fmf{phantom}{ii0,b1,b2} \fmf{phantom}{t1,v1,b1}
       \fmf{phantom}{t2,b2} \fmf{phantom}{t3,b3}
       \fmffreeze
       \fmf{photon}{ii1,v1} \fmf{fermion}{t3,v1} \fmf{fermion,label=$\psi$}{b3,t3} \fmf{fermion}{v1,b3}
       \fmf{photon}{b3,oo1} \fmf{photon}{t3,oo2}
       \fmflabel{$V^{(0)}$}{ii1} \fmflabel{$V^{(1)}$}{oo1} \fmflabel{$V^{(1)}$}{oo2}
      \end{fmfgraph*}
      \end{fmffile}}      ~~~+~~~~~~
      \raisebox{-4ex}[0cm][0cm]{\unitlength=0.5mm
      \begin{fmffile}{GCS_pYY1bb}
      \begin{fmfgraph*}(30,30)
       \fmfpen{thick}
       \fmfleft{i1} \fmfright{o1,o2}
       \fmf{boson}{i1,v1} \fmf{boson}{v1,o1} \fmf{boson}{v1,o2}
       \fmffreeze
       \fmflabel{$V^{(0)}$}{i1}\fmflabel{$V^{(1)}$}{o1}\fmflabel{$V^{(1)}$}{o2}
      \end{fmfgraph*}
      \end{fmffile}} $\Bigg)=0$
      \vskip 2cm
~~~~~~$q^\n ~\Bigg($~~~~
      \raisebox{-4ex}[0cm][0cm]{\unitlength=0.4mm
      \begin{fmffile}{pYYloop222}
      \begin{fmfgraph*}(60,40)
       \fmfpen{thick} \fmfleft{ii0,ii1,ii2} \fmfstraight \fmffreeze \fmftop{ii2,t1,t2,t3,oo2}
       \fmfbottom{ii0,b1,b2,b3,oo1} \fmf{phantom}{ii2,t1,t2} \fmf{phantom}{ii0,b1,b2} \fmf{phantom}{t1,v1,b1}
       \fmf{phantom}{t2,b2} \fmf{phantom}{t3,b3}
       \fmffreeze
       \fmf{photon}{ii1,v1} \fmf{fermion}{t3,v1} \fmf{fermion,label=$\psi$}{b3,t3} \fmf{fermion}{v1,b3}
       \fmf{photon}{b3,oo1} \fmf{photon}{t3,oo2}
       \fmflabel{$V^{(0)}$}{ii1} \fmflabel{$V^{(1)}$}{oo1} \fmflabel{$V^{(1)}$}{oo2}
      \end{fmfgraph*}
      \end{fmffile}}      ~~~+~~~~~~
      \raisebox{-4ex}[0cm][0cm]{\unitlength=0.5mm
      \begin{fmffile}{GCS_pYY1bb}
      \begin{fmfgraph*}(30,30)
       \fmfpen{thick}
       \fmfleft{i1} \fmfright{o1,o2}
       \fmf{boson}{i1,v1} \fmf{boson}{v1,o1} \fmf{boson}{v1,o2}
       \fmffreeze
       \fmflabel{$V^{(0)}$}{i1}\fmflabel{$V^{(1)}$}{o1}\fmflabel{$V^{(1)}$}{o2}
      \end{fmfgraph*}
      \end{fmffile}} $\Bigg)=0$
      \vskip 1cm
\caption{The Ward identities for the amplitude $V^{(0)}_\r(p+q)\to V^{(1)}_\m(p) \, V^{(1)}_\n(q)$ in the unbroken
phase include the GCS as well as the axionic couplings. The solid lines represent fermions and the wiggle lines
are gauge fields. Dashed lines are scalars.
Each depicted diagram also contains the exchange $(\m,p)\leftrightarrow (\n, q)$.} \label{DiagramsUnbroken}
\end{figure}
The total fermionic triangle is given by
     \be
      \D_{\r \m \n}^{011}(p,q;0) =-{1\over{16}} \, \sum_f Q_f (Y_f)^2 \ \G_{\r \m \n}(p,q;0)
                                   =-{\cA^{(1)} \over{16}} \,    \ \G_{\r \m \n}(p,q;0)
     \ee
The superscript indices in the l.h.s. stand for the gauge groups of the vector fields involved in the process.
$\G_{\r \m \n}(p,q;0)$ can be parametrized as in (\ref{Ros}).
For a symmetric distribution of the anomaly (see Appendix \ref{GCSabsorption}), we have
     \bea
      (p+q)^\r  \D_{\r \m \n}^{011}(p,q;0) &=&{1\over3}\frac{ \cA^{(1)} }{32 \pi^2} \e_{\m \n \a \b} p^\a q^\b \nn\\
       p^\m  \D_{\r \m \n}^{011}(p,q;0) &=& {1\over3} \frac{ \cA^{(1)} }{32 \pi^2}  \e_{\n \r \a \b} q^\a p^\b \nn\\
       q^\n  \D_{\r \m \n}^{011}(p,q;0) &=& {1\over3} \frac{ \cA^{(1)} }{32 \pi^2}  \e_{\r \m \a \b} q^\a p^\b
     \eea
Denoting by
\be
 (GS)^{11}_{\m \n}=-2 i b^{(1)}_2 \e_{\m \n \a \b} p^\a q^\b
\ee
the axion interaction vertex and by
\be
(GCS)^{011}_{\r \m \n}=2 d_5 \e_{\r \n \m \a} (p-q)^\a
\ee
 the GCS coupling, the Ward identities in Fig. \ref{DiagramsUnbroken} correspond to
     \bea
      (p+q)^\r \Big(\D_{\r \m \n}^{011}(p,q;0) +(GCS)^{011}_{\r\m\n}\Big)+2i b_3 (GS)^{11}_{\m\n}&=&0\nn \\
      p^\m  \Big(\D_{\r \m \n}^{011}(p,q;0) +(GCS)^{011}_{\r\m\n}\Big) &=&0\nn \\
      q^\n  \Big(\D_{\r \m \n}^{011}(p,q;0) +(GCS)^{011}_{\r\m\n}\Big) &=&0
     \label{WardIdAYY}\eea
They fix the parameters
  \bea
b^{(1)}_2 b_3 = - \frac{\cA^{(1)}}{128 \pi^2} ~~~~~~~~~~~ d_5 = \frac{\cA^{(1)}}{192\pi^2}
\label{b1d5}
  \eea
In the same way, the cancellation of the remaining mixed anomalies gives
  \bea
&&   b^{(0)}_2 b_3 =-\frac{\cA^{(0)}}{384\pi^2}~
                      \qquad \qquad  b^{(2)}_2 b_3 = -\frac{\cA^{(2)}}{64 \pi^2}~
                       \qquad \qquad b^{(4)}_2 b_3 = -\frac{\cA^{(4)}}{128 \pi^2}\nn\\
&&              ~~~~d_4 =- \frac{\cA^{(4)}}{384 \pi^2} ~\qquad \qquad ~~~~
              d_6= \frac{\cA^{(2)}}{96 \pi^2}~
\label{bsds}  \eea
It is worth noting that the GCS coefficients $d_{4,5,6}$ are fully
determined in terms of the $\cA$'s by the Ward identities, while the $b_2^{(a)}$'s depend
only on the free parameter $b_3$, which is related to the mass of
the anomalous $U(1)$.

\subsubsection{Anomaly cancellation in the broken phase}
It is interesting to study the anomaly cancellation procedure in the broken phase. Focusing again onto the
non-vanishing $\cA^{(1)}\neq 0$, the amplitudes that contribute to the cancellation of the anomaly are given in
Fig. \ref{AYYbroken}, where $m_0=Q_{H_u} |v|/2$ and $m_1=|v|/4$ with \mbox{$|v|=\sqrt{v_u^2+v_d^2}$}.
     \begin{figure}[tb]
  \vskip 1.5cm
%
$(p+q)^\r ~\Bigg($~~~~~~~~~~~~~~~
      \raisebox{-4ex}[0cm][0cm]{\unitlength=0.4mm
      \begin{fmffile}{pYYloop3}
      \begin{fmfgraph*}(60,40)
       \fmfpen{thick} \fmfleft{ii0,ii1,ii2} \fmfstraight \fmffreeze \fmftop{ii2,t1,t2,t3,oo2}
       \fmfbottom{ii0,b1,b2,b3,oo1} \fmf{phantom}{ii2,t1,t2} \fmf{phantom}{ii0,b1,b2} \fmf{phantom}{t1,v1,b1}
       \fmf{phantom}{t2,b2} \fmf{phantom}{t3,b3}
       \fmffreeze
       \fmf{photon}{ii1,v1} \fmf{fermion}{t3,v1} \fmf{fermion,label=$\psi$}{b3,t3} \fmf{fermion}{v1,b3}
       \fmf{photon}{b3,oo1} \fmf{photon}{t3,oo2}
       \fmflabel{$V^{(0)}_\r(p+q)$}{ii1}
       \fmflabel{$V^{(1)}_\m(p)$}{oo1}
       \fmflabel{$V^{(1)}_\n(q)$}{oo2}
      \end{fmfgraph*}
      \end{fmffile}}      ~~~~+~~~~~~~~
      \raisebox{-4ex}[0cm][0cm]{\unitlength=0.5mm
      \begin{fmffile}{GCS_pYY1bb}
      \begin{fmfgraph*}(30,30)
       \fmfpen{thick}
       \fmfleft{i1} \fmfright{o1,o2}
       \fmf{boson}{i1,v1} \fmf{boson}{v1,o1} \fmf{boson}{v1,o2}
       \fmffreeze
       \fmflabel{$V^{(0)}$}{i1}\fmflabel{$V^{(1)}$}{o1}\fmflabel{$V^{(1)}$}{o2}
      \end{fmfgraph*}
      \end{fmffile}} $\Bigg)~+$
\vskip 1.8cm
~~~~~~~~~~~~~~~~~~~~~~~~~~~~~~~~~~$+2ib_3~ \Bigg($~
      \raisebox{-4ex}[0cm][0cm]{ \unitlength=0.5mm
      \begin{fmffile}{axionYY}
      \begin{fmfgraph*}(30,30)
      \fmfpen{thick} \fmfleft{i1} \fmfright{o1,o2}
      \fmf{dashes}{i1,v1} \fmf{photon}{v1,o2}
      \fmf{photon}{v1,o1} \fmffreeze \fmflabel{$V^{(1)}$}{o1} \fmflabel{$V^{(1)}$}{o2} \fmflabel{$\f$}{i1}
      \end{fmfgraph*}
      \end{fmffile}} $\Bigg)$
~$+im_0~ \Bigg($~~~~
      \raisebox{-4ex}[0cm][0cm]{\unitlength=0.4mm
      \begin{fmffile}{GoldYYloop1a}
      \begin{fmfgraph*}(60,40)
       \fmfpen{thick} \fmfleft{ii0,ii1,ii2} \fmfstraight \fmffreeze \fmftop{ii2,t1,t2,t3,oo2}
       \fmfbottom{ii0,b1,b2,b3,oo1} \fmf{phantom}{ii2,t1,t2} \fmf{phantom}{ii0,b1,b2} \fmf{phantom}{t1,v1,b1}
       \fmf{phantom}{t2,b2} \fmf{phantom}{t3,b3}
       \fmffreeze
             \fmf{dashes}{ii1,v1}
             \fmf{fermion}{t3,v1} \fmf{fermion,label=$\psi$}{b3,t3} \fmf{fermion}{v1,b3}
       \fmf{photon}{b3,oo1} \fmf{photon}{t3,oo2}
       \fmflabel{$V^{(1)}$}{oo1} \fmflabel{$V^{(1)}$}{oo2} \fmflabel{$NG$}{ii1}
      \end{fmfgraph*}
      \end{fmffile}} ~~$\Bigg)~=~0$
       \vskip 2.2cm
$p^\m \Bigg($~~~~~
      \raisebox{-4.3ex}[0cm][0cm]{\unitlength=0.4mm
      \begin{fmffile}{pYYloop222}
      \begin{fmfgraph*}(60,40)
       \fmfpen{thick} \fmfleft{ii0,ii1,ii2} \fmfstraight \fmffreeze \fmftop{ii2,t1,t2,t3,oo2}
       \fmfbottom{ii0,b1,b2,b3,oo1} \fmf{phantom}{ii2,t1,t2} \fmf{phantom}{ii0,b1,b2} \fmf{phantom}{t1,v1,b1}
       \fmf{phantom}{t2,b2} \fmf{phantom}{t3,b3}
       \fmffreeze
       \fmf{photon}{ii1,v1} \fmf{fermion}{t3,v1} \fmf{fermion,label=$\psi$}{b3,t3} \fmf{fermion}{v1,b3}
       \fmf{photon}{b3,oo1} \fmf{photon}{t3,oo2}
       \fmflabel{$V^{(0)}$}{ii1} \fmflabel{$V^{(1)}$}{oo1} \fmflabel{$V^{(1)}$}{oo2}
      \end{fmfgraph*}
      \end{fmffile}}      ~~+~~~~~~~
      \raisebox{-4.3ex}[0cm][0cm]{\unitlength=0.5mm
      \begin{fmffile}{GCS_pYY1bb}
      \begin{fmfgraph*}(30,30)
       \fmfpen{thick}
       \fmfleft{i1} \fmfright{o1,o2}
       \fmf{boson}{i1,v1} \fmf{boson}{v1,o1} \fmf{boson}{v1,o2}
       \fmffreeze
       \fmflabel{$V^{(0)}$}{i1}\fmflabel{$V^{(1)}$}{o1}\fmflabel{$V^{(1)}$}{o2}
      \end{fmfgraph*}
      \end{fmffile}} $~~\Bigg)$
      $+i m_1~
      \Bigg($~~~~~
      \raisebox{-4.3ex}[0cm][0cm]{\unitlength=0.4mm
      \begin{fmffile}{pYGoldloop1}
      \begin{fmfgraph*}(60,40)
       \fmfpen{thick} \fmfleft{ii0,ii1,ii2} \fmfstraight \fmffreeze \fmftop{ii2,t1,t2,t3,oo2}
       \fmfbottom{ii0,b1,b2,b3,oo1} \fmf{phantom}{ii2,t1,t2} \fmf{phantom}{ii0,b1,b2} \fmf{phantom}{t1,v1,b1}
       \fmf{phantom}{t2,b2} \fmf{phantom}{t3,b3}
       \fmffreeze
       \fmf{photon}{ii1,v1} \fmf{fermion}{t3,v1} \fmf{fermion,label=$\psi$}{b3,t3} \fmf{fermion}{v1,b3}
       \fmf{dashes}{b3,oo1} \fmf{photon}{t3,oo2}
       \fmflabel{$V^{(0)}$}{ii1} \fmflabel{$NG$}{oo1} \fmflabel{$V^{(1)}$}{oo2}
      \end{fmfgraph*}
      \end{fmffile}}  ~~$ \Bigg)  ~=~0$
      \vskip 2.2cm
$q^\n \Bigg($~~~~~
      \raisebox{-4.3ex}[0cm][0cm]{\unitlength=0.4mm
      \begin{fmffile}{pYYloop222}
      \begin{fmfgraph*}(60,40)
       \fmfpen{thick} \fmfleft{ii0,ii1,ii2} \fmfstraight \fmffreeze \fmftop{ii2,t1,t2,t3,oo2}
       \fmfbottom{ii0,b1,b2,b3,oo1} \fmf{phantom}{ii2,t1,t2} \fmf{phantom}{ii0,b1,b2} \fmf{phantom}{t1,v1,b1}
       \fmf{phantom}{t2,b2} \fmf{phantom}{t3,b3}
       \fmffreeze
       \fmf{photon}{ii1,v1} \fmf{fermion}{t3,v1} \fmf{fermion,label=$\psi$}{b3,t3} \fmf{fermion}{v1,b3}
       \fmf{photon}{b3,oo1} \fmf{photon}{t3,oo2}
       \fmflabel{$V^{(0)}$}{ii1} \fmflabel{$V^{(1)}$}{oo1} \fmflabel{$V^{(1)}$}{oo2}
      \end{fmfgraph*}
      \end{fmffile}}      ~~+~~~~~~~
      \raisebox{-4.3ex}[0cm][0cm]{\unitlength=0.5mm
      \begin{fmffile}{GCS_pYY1bb}
      \begin{fmfgraph*}(30,30)
       \fmfpen{thick}
       \fmfleft{i1} \fmfright{o1,o2}
       \fmf{boson}{i1,v1} \fmf{boson}{v1,o1} \fmf{boson}{v1,o2}
       \fmffreeze
       \fmflabel{$V^{(0)}$}{i1}\fmflabel{$V^{(1)}$}{o1}\fmflabel{$V^{(1)}$}{o2}
      \end{fmfgraph*}
      \end{fmffile}}
      $~~\Bigg)$
      $+i m_1~
      \Bigg($~~~~~
      \raisebox{-4.3ex}[0cm][0cm]{\unitlength=0.4mm
      \begin{fmffile}{YpGoldloop1}
      \begin{fmfgraph*}(60,40)
       \fmfpen{thick} \fmfleft{ii0,ii1,ii2} \fmfstraight \fmffreeze \fmftop{ii2,t1,t2,t3,oo2}
       \fmfbottom{ii0,b1,b2,b3,oo1} \fmf{phantom}{ii2,t1,t2} \fmf{phantom}{ii0,b1,b2} \fmf{phantom}{t1,v1,b1}
       \fmf{phantom}{t2,b2} \fmf{phantom}{t3,b3}
       \fmffreeze
       \fmf{photon}{ii1,v1} \fmf{fermion}{t3,v1} \fmf{fermion,label=$\psi$}{b3,t3} \fmf{fermion}{v1,b3}
       \fmf{photon}{b3,oo1} \fmf{dashes}{t3,oo2}
       \fmflabel{$V^{(0)}$}{ii1} \fmflabel{$V^{(1)}$}{oo1} \fmflabel{$NG$}{oo2}
      \end{fmfgraph*}
      \end{fmffile}}  ~~$ \Bigg)  ~=~0$
      \vskip 1cm
\caption{The Ward identities for the amplitude $V^{(0)}_\r(p+q)\to V^{(1)}_\m(p) \, V^{(1)}_\n(q)$ in the broken phase.}
\label{AYYbroken}
\end{figure}
In the broken phase, additional contributions coming from the NG boson exchange must be added.
We denote by $\D_{\r\m\n}(p,q;m_f)$ the modified triangle diagram where also massive fermions circulate in the
loop and by $(NG)_{\r\m\n}$ the triangle diagram with a NG boson on an external leg. Note that
$(GS)_{\r\m\n}$ and $(GCS)_{\r\m\n}$ are the same as in the unbroken phase. The amplitude satisfies again the
usual Ward identities (\ref{WardIdAYY}). In order to clarify the mechanism, we will focus on a single Ward
identity
\be (p+q)^\r \Big(\D_{\r \m \n}^{011}(p,q;m_f) +(GCS)^{011}_{\r\m\n}\Big)+2i b_3 (GS)^{11}_{\m\n} +i m_0 (NG)_{\m
\n}^{11}=0 \ee
From now on the $(p,q;m_f)$ dependence will be explicit only when needed.
Splitting $\D$ and $(NG)$ terms into the sums over SM fermions and higgsinos we obtain
\bea
        \D_{\r \m \n}^{011}&=&\left.    \D_{\r \m \n}^{011} \right|_{SM}+\left.   \D_{\r \m \n}^{011} \right|_{\tilde H_{u,d}}    \\
     (NG)_{\m \n}^{11}&=&\left.  (NG)_{\m \n}^{11}  \right|_{SM}+\left.   (NG)_{\m \n}^{11} \right|_{\tilde H_{u,d}}
     \eea
Since we have
\be
      \left. (p+q)^\r  \D_{\r \m \n}^{011} \right|_{SM}
     =\frac{1}{48 \pi^2} \sum_{f \in SM}
      \[{1\over2} \ t_f^{011} +t_f^{NG11} \ m_f^2 \ I_0 \] \e_{\m \n \a \b} p^\a q^\b
     \label{DeltamassiveSM}
\ee
where the integral $I_0$ is
\be I_0 (p,q;m_f) = -\int_0^1 dx \int_0^{1-x} dy \frac{1}{y (1-y) p^2 +
     x(1-x) q^2 + 2 x y \,p\cdot q - m_f^2} \label{I_0integral}
\ee
     and $t_f^{011}$, $t_f^{NG11}$ are defined in Table \ref{tf's},
  \begin{table}
  \centering
  \begin{tabular}{|c|c|c|}
   \hline                $f$       & $t_f^{011}$                    & $t_f^{NG11}$  \\
   \hline $\n_e$, $\n_\m$, $\n_\t$          & $Q_L Y_L^2$ & $0$   \\
   \hline $e$, $\m$, $\t$          & $Q_L Y_L^2+Q_{E^c} Y_{E^c}^2$ & $Q_{H_d} \( 3 Y_L^2 + 3 Y_L Y_{H_d} + Y_{H_d}^2 \)$   \\
   \hline $u$, $c$, $t$            & $N_c\(Q_Q Y_Q^2+Q_{U^c} Y_{U^c}^2\)$ & $N_c \,Q_{H_u} \( 3 Y_Q^2 + 3 Y_Q Y_{H_u} + Y_{H_u}^2 \)$  \\
   \hline $d$, $s$, $b$            & $N_c\(Q_Q Y_Q^2+Q_{D^c} Y_{D^c}^2\)$ & $N_c \, Q_{H_d} \( 3 Y_Q^2 + 3 Y_Q Y_{H_d} + Y_{H_d}^2 \)$ \\
   \hline
  \end{tabular}
  \caption{Definition of $t_f^{011}$ and $t_f^{NG11}$, where $N_c=3$ is the number of colours.} \label{tf's}
  \end{table}
the Ward identity of the SM fermionic loop has a new contribution due to the masses of the fermions. Similarly, for the corresponding $NG$ term we get
\be i m_0 \left. (NG)_{\m \n}^{11}  \right|_{SM} = - \frac{1}{48 \pi^2} \sum_{f \in SM}
\[t_f^{NG11} \ m_f^2 \ I_0\] \e_{\m \n \a \b} p^\a q^\b
\label{NGpartSM}
\ee
Summing (\ref{DeltamassiveSM}, \ref{NGpartSM}), the massive contribution in the fermionic loop is exactly cancelled by the NG ones, giving
\bea
\[(p+q)^\r  \D_{\r \m \n}^{011}(p,q;m_f) + i m_0 (NG)_{\m \n}^{11} \]_{SM}&=&\frac{1}{96 \pi^2} \sum_{f \in SM}
                   t_f^{011} \e_{\m \n \a \b} p^\a q^\b \nn\\
&=&  (p+q)^\r  \left. \D_{\r \m \n}^{011}(p,q;0) \right|_{SM}  \label{WImassiveSM} \eea
The contribution of the diagrams involving the higgsinos vanishes
      \bea
      \[(p+q)^\r  \D_{\r \m \n}^{011}(p,q;m_f) + i m_0 (NG)_{\m \n}^{11} \]_{\tilde H_{u,d}}&=&\frac{1}{96 \pi^2} \sum_{f \in \tilde H_{u,d}} Q_f Y_f^2 \e_{\m \n \a \b} p^\a q^\b=0~~~~~~~~
     \label{WImassiveSusy}
     \eea
Summing (\ref{WImassiveSM}, \ref{WImassiveSusy}) we get
\be
      \[(p+q)^\r  \D_{\r \m \n}^{011}(p,q;m_f) + i m_0 (NG)_{\m \n}^{11} \] = \frac{\cA^{(1)}}{96 \pi^2} \e_{\m \n \a \b} p^\a q^\b  =    (p+q)^\r  \ \D_{\r \m \n}^{011}(p,q;0) \label{Gbrokenunbroken}
\ee
Thus the contribution to the Ward Identities of the triangle diagrams is exactly the same as in the unbroken phase.

\subsection{Soft breaking terms}
The total soft breaking lagrangian can be written as
   \be
    \L_{soft}=\L_{soft}^{MSSM}+\L_{soft}^{new}
   \ee
with
   \bea
    \L_{soft}^{MSSM} &=& - {1\over2}  \sum_{a=1}^3 \(M_a \l^{(a)} \l^{(a)} + h.c. \) -
                 \( m^2_{Q_{ij}} \Qt_i \Qt_j^\dag + m^2_{U_{ij}} \Ut_i \Uts_j + m^2_{D_{ij}} \Dt_i \Dts_j  \right. \nn\\
                    &&\left.+m^2_{L_{ij}} \Lt_i \Lt_j^\dag + m^2_{E_{ij}} \Et_i \Ets_j + m^2_{h_u} |h_u|^2 + m^2_{h_d} |h_d|^2 \) \nn\\
                    &&-\( a_u^{ij} \Qt_i \Ut_j h_u - a_d^{ij} \Qt_i \Dt_j h_d - a_e^{ij} \Lt_i \Et_j h_d + b h_u h_d + h.c. \)\label{LsoftStandard}
   \eea
   and
   \bea \L_{soft}^{new}=- {1\over2}  \(M_0 \l^{(0)} \l^{(0)} + h.c. \) - {1\over2}  \(M_S \psi_S \psi_S  + h.c. \)
   \eea
where $\l^{(0)}$ is the gaugino of the added $U(1)'$ and $\psi_S$ is the axino.
We allow a soft mass term for the axino since it couples only through GS interactions and not through Yukawa
interactions \cite{Girardello:1981wz}. Notice also that a mass term for the axion $\f$ is not allowed since it
transforms non trivially under the anomalous $U(1)'$ gauge transformation (\ref{U1Trans}).

\section{Model setup}\label{Consequences}

In this Section we analyze the effects of the additional terms
on the rest of the lagrangian.

\subsection{Kinetic diagonalization of U(1)'s}

As we mentioned before, the St\"uckelberg multiplet contains a complex scalar field whose real part gets an
expectation value that modifies the coupling constant (\ref{couplingconst}). Therefore, the second line in
(\ref{Laxion}) contributes to the kinetic terms for the gauge fields and the term $\langle \a\rangle
W^{(1)}W^{(0)}$ gives a kinetic mixing between the $V^{(1)}$ and $V^{(0)}$ gauge bosons. Redefining as usual
$V^{(0)} \to 2 g_0 V^{(0)}$, $V^{(1)} \to 2 g_1 V^{(1)}$ we get
\be
  \left. \( {1\over4}  W^{(0)} W^{(0)} + {1\over4} W^{(1)} W^{(1)}   +{\d\over2} W^{(1)} W^{(0)} \) \right|_{\th^2}
\label{kinmixlag}
\ee
with $\d = - 4 b^{(4)}_2 g_0 g_1  \langle \a\rangle $. In order to diagonalize the kinetic terms, we use the
matrix
  \be
      \( \begin{array}{c} V^{(0)}\\
                          V^{(1)} \end{array} \) = \( \begin{array}{cc} C_\d &  0 \\
                         -S_\d  &  1 \end{array} \)
            \( \begin{array}{c} V_C\\
                                V_B \end{array} \)
 \label{cdelta} \ee
  where $C_\d = 1/\sqrt{1-\d^2}$ and $S_\d =  \d C_\d$.
Let us stress that in this case the mixing is a consequence
of the anomaly cancellation procedure. Note that, since $b_2^{(4)}\sim b_3^{-1}\sim M_{V^{(0)}}^{-1}$
(see eq. (\ref{bsds})), where $M_{V^{(0)}}$ is the mass of the anomalous $U(1)$ that we assume to be in the TeV range,
this mixing is tiny and can be ignored for our purposes.

\subsection{D and F terms}

The additional fields give rise also to D and F  terms. More precisely, D term
contributions come from: (i) the kinetic terms of chiral multiplets and (ii) the axionic lagrangian, providing
(\ref{Laxion})
   \bea
    \L_D&=&\frac{1}{2}\sum_{a=0}^3 D^{(a)}_{k_a} D^{(a)}_{k_a} + \sum_{a=0}^3 g_a D^{(a)}_{k_a} z_i^\dag (T^{(a)}_{k_a})_j^i z^j
    + 4 g_0 b_3 \langle\a\rangle D^{(0)} + \d D^{(1)} D^{(0)}+\nn\\
                 &&+2 \[\sum_{a=0}^2 g_a^2 \ b^{(a)}_2  \sqrt2 \psi_S \Tr \(  \l^{(a)} D^{(a)} \)
                + g_0 g_1 {b^{(4)}_2\over\sqrt2} \psi_S \( \l^{(1)} D^{(0)} + \l^{(0)} D^{(1)} \) +h.c. \]\nn\\
    \eea
where $a=0,1,2,3$ denotes, as usual, the gauge group factors, $z_i$ are the lowest components of the $i$-th chiral multiplet (except the multiplet
which contains the axion) and
$T^{(a)}_{k_a}$, $k_a=1,\ldots,{\rm dim G}^{(a)}$, are the generators of the corresponding gauge groups, ${\rm G}^{(a)}$.
Solving the equations of motion for the D's and substituting back we obtain
   \bea
    \L_{D_C} ~&=&- {1\over2} \left\{   \[C_\d g_0 \sum_f Q_f |z_f|^2 - S_\d g_1 \sum_f Y_f |z_f|^2 \] \right.
                  + C_\d 4 g_0 b_3 \langle\a\rangle \nn\\
                  && ~~~~~~ +2\sqrt2 b^{(0)}_2 g_0^2 \[ \psi_S \( C_\d^2 \l_C\)+h.c. \]
                  +2\sqrt2 b^{(1)}_2 g_1^2 \[ \psi_S \( S_\d^2\l_C -S_\d \l_B \)+h.c. \]\nn\\
                  &&~~~~~~+\sqrt2 b^{(4)}_2 g_0 g_1 \[ \psi_S \( C_\d \l_B - 2 C_\d S_\d \l_C\)+h.c. \]\Bigg\}^2\label{dcterm}\\
\L_{D_B} ~&=&- {1\over2} \Bigg\{ g_1 \sum_f Y_f |z_f|^2
                +2\sqrt2 b^{(1)}_2 g_1^2 \[ \psi_S \( \l_B -S_\d \l_C  \)+h.c. \]+\nn\\
               &&~~~~~~+\sqrt2 b^{(4)}_2 g_0 g_1 \[  \psi_S  C_\d  \l_C +h.c. \]\Bigg\}^2\label{dbterm}\\
\L_{D^{(2)}} &=& -\frac{1}{2} \sum_k\left\{g_2 z_i^\dag (T^{(2)}_k)_j^i z^j
          + b^{(2)}_2 g_2^2 \[  \sqrt2 \psi_S  \l^{(2)}_k  + h.c. \]\right\}^2\label{d2term}\\
      \L_{D^{(3)}} &=& -\frac{1}{2} \sum_k \left\{ g_3 z_i^\dag (T^{(3)}_k)_j^i z^j
          \right\}^2\label{d3term}
     \eea
Similarly, the F term contributions are
  \bea
   \L_F &=& \sum_{f \in MSSM} \( F^f F_f^\dag-\frac{\pd W}{\pd z^f} F^f-\frac{\pd W^\dag}{\pd z^\dag_f} F^\dag_f\)\nn\\
   &&+\frac{1}{2} F_S F_S^\dag +
   \frac{1}{2} \left\{F_S \[ \sum_{a=0}^2 b^{(a)}_2 \Tr \(  \l^{(a)} \l^{(a)} \) + b^{(4)}_2 \l^{(1)} \l^{(0)} \] +h.c. \right\}
  \eea
where the first line is the standard MSSM F term contribution
while the second line contains the new axionic terms. Solving the
EOM, and rescaling $V\to 2g V$ we get
  \bea
   \L_{F_S} &=& -8 \[ \sum_a b^{(a)}_2  g_a^2 \Tr \(  \l^{(a)} \l^{(a)} \)+ g_1 g_0  b^{(4)}_2 \l^{(1)} \l^{(0)}\]\nn\\
      &&~~~ \times \[ \sum_a b^{(a)}_2 g_a^2 \Tr \(  \lb^{(a)} \lb^{(a)} \) + g_1 g_0 b^{(4)}_2 \lb^{(1)} \lb^{(0)}\]
\label{fsterm}
  \eea
Eq. (\ref{fsterm}) can also be written in the basis (\ref{cdelta}), but we will not need this term in the following.

We would like to mention that no D and F terms are coming from
the GCS since they include only vector multiplets in an
antisymmetric form. Our results are in accordance with
\cite{DeRydt:2007vg}.

 \subsection{Scalar potential}

As we have seen in the previous section, the additional F terms
(\ref{fsterm}) do not give any contribution to the scalar
potential. The D$_B$, D$^{(2)}$ and D$^{(3)}$ terms (see eq.
(\ref{dbterm}), (\ref{d2term}) and (\ref{d3term})) provide the
usual contributions to the MSSM potential. The only new
contribution comes from the first line of (\ref{dcterm}). Thus the
scalar potential can be written as
  \bea
   V&=&V_{MSSM}+V_{D_C} \label{V}\\
   V_{D_C} &=& {1\over2} \left\{   \[C_\d g_0 \sum_f Q_f |z_f|^2 - S_\d g_1 \sum_f Y_f |z_f|^2 \]
                  + C_\d 4 g_0 b_3 \langle\a\rangle \right\}^2
  \eea
Solving the equations for the minima of the potential
  \be
   \frac{\pd V}{\pd z_f} =0
  \ee
we get $\la z_f \ra =0$ for all the sfermions as in the
 MSSM case. Inserting back these vevs into (\ref{V}) we get the following Higgs scalar potential
  \bea
   V_h &=& \Big\{ |\m|^2+m^2_{h_u} + 4 g_0^2 b_3 \langle\a\rangle C_\d X_\d \Big\} \Big(|h_u^0|^2 + |h_u^+|^2\Big) \nn\\
          &&+\Big\{ |\m|^2+m^2_{h_d} - 4 g_0^2 b_3 \langle\a\rangle C_\d  X_\d \Big\} \Big(|h_d^0|^2 + |h_d^-|^2\Big)\nn\\
       && + \Big\{\frac{1}{2} \( g_0 X_\d \)^2 +\frac{1}{8}(g_1^2+g_2^2)\Big\} \Big(|h_u^0|^2 +|h_u^+|^2 - |h_d^0|^2 - |h_d^-|^2 \Big)^2\nn\\
       &&+\Big\{ b\, (h_u^+ h_d^- - h_u^0 h_d^0) + h. c.\Big\}    + \half g_2^2 |h_u^+ h_d^{0*} + h_u^0 h_d^{-*}|^2
  \eea
which can be brought to the same form of the MSSM potential, after the
following redefinitions
  \bea
    m^2_{h_u} + 4 g_0^2 b_3 \langle\a\rangle C_\d X_\d &\to& \tilde m^2_{h_u} \nn\\
    m^2_{h_d} - 4 g_0^2 b_3 \langle\a\rangle C_\d X_\d &\to& \tilde m^2_{h_d} \nn\\
   \(\( g_0 X_\d \)^2 +\frac{1}{4}(g_1^2+g_2^2) \)v^2&\to& \tilde m^2_Z \label{tildeMs}
  \eea
  where
  \be
   g_0 X_\d= C_\d g_0 \QHu - \half S_\d g_1\label{xdelta}
  \ee
At the minimum, we recover the MSSM result $\la h_u^+ \ra=\la h_d^-
\ra=0$ for the Higgs charged components. Defining $\la h^0_i \ra= v_i/\sqrt2$ , $v_u^2+v_d^2 = v^2$ and $v_u/v_d=\tan\b$ we can
still write the tree level conditions for the electroweak symmetry
breaking as
  \bea
   b^2 &>& \( |\m|^2+ \tilde m^2_{h_u} \) \( |\m|^2+ \tilde m^2_{h_d} \) \\
   2 b &<&  2 |\m|^2+ \tilde m^2_{h_u} + \tilde m^2_{h_d}
  \eea
in complete analogy with the MSSM case (using $\tilde{m}$'s).

\subsection{Higgs sector}

It is worth noting that in our model there is no axi-higgs mixing.
This is due to the fact that we do not consider scalar potential terms
for the axion  (on the contrary to \cite{Coriano':2005js}).

After the electroweak symmetry breaking we have four gauge
generators that are broken, so we have four longitudinal degrees of freedom. One of them
is the axion, while the other three are the usual NG bosons coming
from the Higgs sector.

As it was mentioned above, the potential has the standard MSSM
form, upon the redefinitions (\ref{tildeMs}). The Higgs scalar
fields consist of two complex $SU(2)_L$-doublets, or eight real,
scalar degrees of freedom. When the electroweak symmetry is
broken, three of them are the would-be NG bosons
$G^0$, $G^\pm$. The remaining five Higgs scalar mass eigenstates
consist of two CP-even neutral scalars $h^0$ and $H^0$, one CP-odd
neutral scalar $A^0$ and a charge $+1$ scalar $H^+$ as well as its
charge conjugate $H^-$ with charge $-1$.\footnote{
We define $G^{-} = G^{+*}$ and $H^- = H^{+*}$. Also, by convention, $h^0$ is lighter
than $H^0$.} The gauge-eigenstate fields can be expressed in terms
of the mass eigenstate fields as
    \bea
     \( \begin{array}{c} h_u^0 \\
                         h_d^0 \end{array} \) &=& {1\over \sqrt{2}} \(\begin{array}{c} v_u \\
                                                                           v_d \end{array}\) +
                                          {1\over \sqrt{2}} R_\alpha \(\begin{array}{c} h^0 \\
                                                                                      H^0 \end{array}\) +
                                          {i\over \sqrt{2}} R_{\beta_0}\(\begin{array}{c} G^0 \\
                                                                                        A^0 \end{array}\)\label{HiggsMin}\\
     \( \begin{array}{c} h_u^+ \\
                         h_d^{-*} \end{array} \) &=&  R_{\beta_\pm}\(\begin{array}{c} G^+ \\
                                                                                        H^+ \end{array}\)
    \eea
where the orthogonal rotation matrices $R_\alpha, R_{\beta_0}, R_{\beta_\pm}$ are the same as in \cite{Martin:1997ns}
Acting with these matrices on the gauge eigenstate fields we obtain the diagonal mass terms.
Expanding around the minimum (\ref{HiggsMin}) one finds that
$\beta_0 = \beta_\pm = \beta$, and replacing the tilde parameters
(\ref{tildeMs}) we obtain the masses
    \bea
     m_{A^0}^2 &=& 2|\m|^2 + m^2_{h_u} + m^2_{h_d}\\
     m^2_{h^0, H^0} &=& \frac{1}{2} \Bigg\{ m^2_{A^0} +
     \(\( g_0 X_\d \)^2 +\frac{1}{4}(g_1^2+g_2^2) \)v^2 \nn\\
              &&\mp \[\(m_{A^0}^2
              - \(\( g_0 X_\d \)^2 +\frac{1}{4}(g_1^2+g_2^2) \)v^2\)^2 \right. \nn\\
     &&\left. + 4 \(\( g_0 X_\d \)^2 +\frac{1}{4}(g_1^2+g_2^2) \)v^2
     m_{A^0}^2  \sin^2 (2\beta)\]^\half \Bigg\}\>\>\>\>\>{} \label{eq:m2hH}\\
     m^2_{H^\pm} &=& m^2_{A^0} + m_W^2 =  m^2_{A^0} +  g_2^2 \frac{v^2}{4} \label{eq:m2Hpm}
    \eea
and the mixing angles
    \bea
    {\sin 2\alpha\over \sin 2 \beta} &=& -{m_{H^0}^2 + m_{h^0}^2 \over m_{H^0}^2 - m_{h^0}^2}  \nn\\
     \quad {\tan 2\a\over \tan 2 \b} &=&
           {m_{A^0}^2+
           \(\( g_0 X_\d \)^2 +\frac{1}{4}(g_1^2+g_2^2) \)v^2 \over  m_{A^0}^2-
     \(\( g_0 X_\d \)^2 +\frac{1}{4}(g_1^2+g_2^2) \)v^2}
    \eea
Notice that only the $h^0$ and $H^0$ masses get modified with
respect to the MSSM, due to the additional anomalous $U(1)'$.

\subsection{Neutral Vectors}   \label{vectmasssol}

There are two mass-sources for the gauge bosons: (i) the
St\"uckelberg mechanism and (ii) the Higgs mechanism. In this extension of
the MSSM, the mass terms for the gauge fields are given by
     \be
      \L_M = \frac{1}{2} \(C_\m \ B_\m \ V^{(2)}_{3\m} \) M^2
                          \( \begin{array}{c} C^\m\\ B^\m\\ V^{(2)\m}_{3} \end{array} \)
     \ee
$C_\mu ,\, B_\mu$ are the lowest components of the vector multiplets $V_C,\,V_B$.
The gauge boson mass matrix is
     \be
      M^2= \( \begin{array}{ccc} M^2_C & ~~~g_0 g_1 \frac{v^2}{2} X_\d& ~~~-g_0 g_2 \frac{v^2}{2} X_\d  \\
                 ... & g_1^2 \frac{v^2}{4} & -g_1 g_2 \frac{v^2}{4}   \\
                 ... & ...  & g_2^2 \frac{v^2}{4}  \\\end{array} \)
  \label{BosonMasses} \ee
where $M^2_C =16 g_0^2 b_3^2 C_\d^2 + g_0^2 (v^2) X_\d^2$ and the lower dots denote the obvious terms under
symmetrization. After diagonalization, we obtain the
eigenstates
 \bea
   A_\m &=&\frac{g_2 B_\m + g_1 V^{(2)}_{3\m}}{\sqrt{g_1^2+g_2^2}}  \label{photon}\\
    Z_{0\m} &=& \frac{g_2 V^{(2)}_{3\m} - g_1 B_\m}{\sqrt{g_1^2+g_2^2}}+g_0 \QHu\frac{\sqrt{g_1^2+g_2^2}  v^2}{2 M_{V^{(0)}}^2} C_\m+{\cal O}[g_0^3,M_{V^{(0)}}^{-3}]  \label{Z0}\\
Z'_\m  &=& C_\m +\frac{g_0 \QHu v^2}{2 M_{V^{(0)}}^2}\(g_1 B_\m- g_2 V^{(2)}_{3\m}\) +{\cal O}[g_0^3,M_{V^{(0)}}^{-3}]
\label{Zprime}
    \eea
and the corresponding masses
   \bea
   M^2_{\g}&=&0\\
    M^2_{Z_0} &=&\frac{1}{4} \(g_1^2+g_2^2\) v^2
                 -(\QHu)^2\frac{\(g_1^2+g_2^2\) g_0^2  v^4}{4 M_{V^{(0)}}^2}+{\cal O}[g_0^3,M_{V^{(0)}}^{-3}]
\label{Z0mass}\\
M^2_{Z'}  &=&M_{V^{(0)}}^2+g_0^2 \[(\QHu)^2 \(1+\frac{g_1^2 v^2+g_2^2
v^2}{4M_{V^{(0)}}^2}\)-\frac{\langle\a\rangle g_1^3 \cA^{(4)}}{64 \pi ^2M_{V^{(0)}}}\] v^2+{\cal
O}[g_0^3,M_{V^{(0)}}^{-3}]~~~~~~~~~
  \label{Zpmass} \eea
where $M_{V^{(0)}}=4 b_3 g_0$ is the mass parameter for the anomalous $U(1)$ and it is assumed to be in the TeV
range. Due to their complicated form, the eigenstates and eigenvalues of $M^2$ (\ref{BosonMasses}) are expressed
as power expansions in  $g_0$ and $1/M_{V^{(0)}}$ keeping only the leading terms. Higher terms are denoted
by ${\cal O}[g_0^3,M_{V^{(0)}}^{-3}]$.

The first eigenstate (\ref{photon}) corresponds to the photon and it is exact to all orders. It slightly
differs from the usual MSSM expression due to the kinetic mixing between $V^{(0)}$ and $V^{(1)}$.

For the rest of the paper, we neglect the kinetic mixing contribution since they are higher loop
effects which go beyond the scope of the present paper.
Then the rotation matrix from the hypercharge to the photon
basis, up to ${\cal O}[g_0^3,M_{V^{(0)}}^{-3}]$  is
    \bea
     \( \begin{array}{c} Z'_\m\\
                         Z_{0 \m}\\
                         A_\m \end{array} \)
       &=&O_{ij}
     \( \begin{array}{c} V^{(0)}_\m\\
                         V^{(1)}_\m\\
                         V^{(2)}_{3\m} \end{array} \)\label{Oij}\\
      &=&\( \begin{array}{ccc}      1& g_1 \frac{g_0 \QHu v^2}{2 M_{V^{(0)}}^2}  ~~& -g_2 \frac{g_0 \QHu v^2}{2 M_{V^{(0)}}^2}  \\
                                g_0 \QHu \frac{\sqrt{g_1^2+g_2^2}  v^2}{2M_{V^{(0)}}^2} ~~~& - \frac{g_1}{{\sqrt{g_1^2+g_2^2}}}
& \frac{g_2}{{\sqrt{g_1^2+g_2^2}}}  \\
                                0 & \frac{g_2}{{\sqrt{g_1^2+g_2^2}}} & \frac{g_1}{{\sqrt{g_1^2+g_2^2}}}  \\ \end{array} \)
     \( \begin{array}{c} V^{(0)}_\m\\
                         V^{(1)}_\m\\
                         V^{(2)}_{3\m} \end{array} \) \nn \eea
where $i,j=0,1,2$.

\subsection{Sfermions}

In general, the contributions to the sfermion masses are coming
from (i) the D and F terms in the superpotential and (ii) the
soft-terms. However, in our case, the new contribution comes only
from the $D_C$ terms
    \bea
     V^{D_C}_\text{mass} = \bigg\{ \( C_\d g_0 \QHu + \half S_\d g_1 \) \( \frac{v_u^2-v_d^2}{2}\) + 4 C_\d g_0 b_3 \langle\a\rangle  \bigg\}
                 \bigg\{ \sum_f  \( C_\d g_0  Q_f - S_\d g_1 Y_f \) |y_f|^2 \bigg\} \nn\\
    \eea
where the $y_f$ stand for all possible sfermions.

\subsection{Neutralinos \label{Neutralinos}}

With respect to the MSSM, now we have two new fields: $\psi_S$ and $\l^{(0)}$. Thus, we have
    \be
    \L_{\mbox{neutralino mass}} = -\frac{1}{2} (\psi^{0})^T {\bf M}_{\tilde N} \psi^0 + h. c.
    \ee
where
   \be(\psi^{0})^T= (\psi_S, \ \l_C,\ \l_B,\ \l^{(2)},\ \tilde h_d^0,\ \tilde h_u^0) \label{neutrbase}
    \ee
The neutralino mass matrix $ {\bf M}_{\tilde N} $ gets contributions from (i) the MSSM terms, (ii) the $h-\tilde
h- \l^{(0)}$ couplings, (iii) the new soft-breaking terms $\L^{new}_{soft}$, (iv) the St\"uckelberg action and (v)
the D terms. Finally, we obtain the symmetric matrix
   \be
    {\bf M}_{\tilde N}
     =   \(\begin{array}{cccccc}
           M_S~~~ & m_{SC} & m_{SB} & {2\over\sqrt2} g_2^3 b_2^{(2)} \, \Delta v^2 & 0 & 0 \\
           \dots & M_0 C_\d^2+M_1 S_\d^2~~~ &  -M_1 S_\d  & 0 & - g_0 v_d X_\d  ~~~& g_0 v_u X_\d   \\
           \dots & \dots & M_1 & 0 & -\frac{g_1 v_d}{2} & \frac{g_1 v_u}{2} \\
           \dots & \dots& \dots & M_2 & \frac{g_2 v_d}{2} & -\frac{g_2 v_u}{2} \\
           \dots & \dots & \dots & \dots & 0 & -\m  \\
          \dots & \dots & \dots & \dots & \dots & 0
         \end{array}\) \label{massmatrix} ~~~~
   \ee
where $M_1,~M_2$ are the gaugino masses coming from the soft
breaking terms (\ref{LsoftStandard}), and
       \bea
    m_{SC}&=&\sqrt2\Bigg\{2\(C_\d^2 g_0^2 b_2^{(0)}+S_\d^2 g_1^2 b_2^{(1)}- C_\d S_\d g_0 g_1 b_2^{(4)} \)
                   \( g_0 X_\d \, \Delta v^2 +  C_\d M_{V^{(0)}} \la \a \ra\)\nn\\
         &&+\half \(-2S_\d g_1^2 b_2^{(1)}+C_\d g_0 g_1 b_2^{(4)} \)g_1 \, \Delta v^2+  \frac{C_\d}{2}M_{V^{(0)}}\Bigg\}\label{msc}\\
     m_{SB} &=& \sqrt2\left\{ \( C_\d g_0 g_1 b_2^{(4)}-2 S_\d g_1^2 b_2^{(1)} \)
                \(g_0 X_\d \, \Delta v^2  + C_\d M_{V^{(0)}} \la \a \ra\)
               + b_2^{(1)} g_1^3 \, \Delta v^2\right\}\nn
   \eea
with $\Delta v^2=v_u^2-v_d^2~$.
It is worth noting that the D terms and kinetic mixing terms are only
higher order corrections and they can be neglected in the computations
of the eigenvalues and eigenstates.

\section{Phenomenology}\label{decays}

In this Section we compute the amplitudes for the decays $Z'
\to Z_0 \, \g$ and $Z' \to Z_0 \, Z_0$\footnote{We would like to
acknowledge T. Tomaras for discussions on this point.} focusing for simplicity on the case $\QHu=0$. In this case there is no mixing between the $V^{(0)}$ and the other
SM gauge fields therefore $Z'=V^{(0)}$ (see (\ref{Oij})).  Notice also that neutralino and chargino contributions to the fermionic triangles identically vanish, giving the same results, for what the decays of interest are concerned, of non-SUSY models.  In Table \ref{couplingsTableQH0} we list all the couplings of the SM fermions with the neutral gauge bosons
  \begin{table}
  \centering
  \begin{tabular}[h]{|c|c|c|c|c|c|}
   \hline                          & $q_f$  &  $v_f^{Z_0}$                & $a_f^{Z_0}$ & $v_f^{Z'}$ & $a_f^{Z'}$ \\
   \hline $\n_e$, $\n_\m$, $\n_\t$ & $0$    &  $1/2$                      &   $1/2$  & $Q_L$ & $Q_L$\\
   \hline $e$, $\m$, $\t$          & $-1$   &  $-1/2+2 \sin^2 \theta_W$   &  $-1/2$  & $2 Q_L$ & $0$\\
   \hline $u$, $c$, $t$            & $2/3$  &  $1/2-4/3 \sin^2 \theta_W$  &   $1/2$  & $2 Q_Q$ & $0$\\
   \hline $d$, $s$, $b$            & $-1/3$ &  $-1/2+2/3 \sin^2 \theta_W$ &  $-1/2$  & $2 Q_Q$ & $0$\\
   \hline
  \end{tabular}
  \caption{Couplings of the SM fermions with the neutral gauge bosons.}\label{couplingsTableQH0}
  \end{table}
where $q_f$ denote the electric charges, $v_f^{Z_0}$ and $a_f^{Z_0}$ are
the vectorial and axial couplings with $Z_0$ and
$v_f^{Z'}$ and $a_f^{Z'}$ are the vectorial and axial couplings with $Z'$,
respectively (see also (\ref{neutralcurrents})).

\subsection{$Z' \to Z_0 \ \g$}

We compute all the relevant diagrams in the $R_\xi$ gauge, thus
removing the interaction vertex $V^\m \partial_\m G_V$ that
involves the massive gauge bosons and the St\"uckelberg or
NG boson. Therefore, the only diagrams that remain are the
fermionic loop, the GCS vertex and a not anomalous remnant
contribution (Fig. \ref{diagzp}). It is possible to show that
the last blob-diagram, that involves several diagrams, is equal to
zero. For the interested reader we give further details in
Appendix~\ref{notanomappdx}.
     \begin{figure}[tb]
     \vskip 1.5cm
      \centering
      \raisebox{-5.2ex}[0cm][0cm]{\unitlength=0.7mm
      \begin{fmffile}{BBBMSSM_ZZgNEW}
      \begin{fmfgraph*}(30,30)
       \fmfpen{thick} \fmfleft{o1} \fmfright{i1,i2} \fmf{boson}{i1,v1,i2}\fmf{boson}{o1,v1}
       \fmfv{decor.shape=circle,decor.filled=hatched, decor.size=.30w}{v1}
       \fmflabel{$Z'^\r(p+q)$}{o1}\fmflabel{$Z_0^\m(p)$}{i1}\fmflabel{$\g^\n(q)$}{i2}
      \end{fmfgraph*}
      \end{fmffile}}
       ~~~=~~~~~~~~~~~~~~~~~~~~~~~~~~~~~~~~~~~~~~~~~~~~~~~~~~~~~~
 \vskip2.5cm
      \raisebox{-4.3ex}[0cm][0cm]{\unitlength=0.4mm
      \begin{fmffile}{ZZg_131}
      \begin{fmfgraph*}(60,40)
       \fmfpen{thick} \fmfleft{ii0,ii1,ii2} \fmfstraight \fmffreeze \fmftop{ii2,t1,t2,t3,oo2}
       \fmfbottom{ii0,b1,b2,b3,oo1} \fmf{phantom}{ii2,t1,t2} \fmf{phantom}{ii0,b1,b2} \fmf{phantom}{t1,v1,b1}
       \fmf{phantom}{t2,b2} \fmf{phantom}{t3,b3}
       \fmffreeze
       \fmf{photon}{ii1,v1} \fmf{fermion}{t3,v1} \fmf{fermion,label=$\psi$}{b3,t3} \fmf{fermion}{v1,b3}
       \fmf{photon}{b3,oo1} \fmf{photon}{t3,oo2}
       \fmflabel{$Z'$}{ii1} \fmflabel{$Z_0$}{oo1} \fmflabel{$\gamma$}{oo2}
      \end{fmfgraph*}
      \end{fmffile}}
      ~~~~~+~~~~
      \raisebox{-5.2ex}[0cm][0cm]{\unitlength=0.7mm
      \begin{fmffile}{GCS_ZZg}
      \begin{fmfgraph*}(30,30)
       \fmfpen{thick}
       \fmfleft{i1} \fmfright{o1,o2}
       \fmf{boson}{i1,v1} \fmf{boson}{v1,o1} \fmf{boson}{v1,o2}
       \fmffreeze
       \fmflabel{$Z'$}{i1}\fmflabel{$Z_0$}{o1}\fmflabel{$\g$}{o2}
      \end{fmfgraph*}
      \end{fmffile}}
      ~~~~~+~~~~
      \raisebox{-5.2ex}[0cm][0cm]{\unitlength=0.7mm
      \begin{fmffile}{BBBMSSM_NAZZg}
      \begin{fmfgraph*}(30,30)
       \fmfpen{thick} \fmfleft{o1} \fmfright{i1,i2} \fmf{boson}{i1,v1,i2}\fmf{boson}{o1,v1}
       \fmfv{decor.shape=circle,decor.filled= shaded, decor.size=.30w}{v1}
       \fmflabel{$Z'$}{o1}\fmflabel{$Z_0$}{i1}\fmflabel{$\g$}{i2}\fmflabel{~~Not An.}{v1}
      \end{fmfgraph*}
      \end{fmffile}}
       \vskip 1.5cm
       \caption{Diagrams for $Z' \to Z_0 \ \g$.} \label{diagzp}
     \end{figure}

    The decay rate for the process is given by
    \be
     \G \(Z' \to Z_0 \g \) = \frac{p_F}{32 \pi^2 \MZp^2} \int |A_\text{TOT}|^2 d\Omega
    \label{decay1}\ee
    where $A_\text{TOT}$ is the total scalar amplitude and $p_F$ is the momentum of the outgoing vectors in
    the CM frame
    \be
     p_F = \frac{\MZp}{2} \left( 1 - \frac{\MZO^2}{\MZp^2} \right)
    \label{decay11}\ee
The square of the total scalar amplitude is given by
    \be
     |A_\text{TOT}|^2 = \frac{1}{3} \sum_{\l'} \e^{\r_1}_{(\l')} \e^{* \r_2}_{(\l')} \
                       \sum_{\l^0} \e^{\n_1}_{(\l^0)} \e^{* \n_2}_{(\l^0)} \
                      \sum_{\l_\g} \e^{\m_1}_{(\l^\g)} \e^{* \m_2}_{(\l^\g)} \
                      A_{\r_1 \m_1 \n_1}^{Z' Z_0 \g} A_{\r_2 \m_2 \n_2}^{* \, Z' Z_0 \g}
    \ee
where $\e$ are the polarizations of the gauge bosons, and $A_{\r
\m \n}$ is the Feynman amplitude of the process. The factor $1/3$
comes from the average over the $Z'$ helicity states. The
polarizations obey to the following completeness relations
    \bea
     \sum_{\l'} \e^{\r_1}_{(\l')} \e^{* \r_2}_{(\l')} &=& - \eta^{\r_1 \r_2} + \frac{k^{\r_1}_{(\l')} k^{\r_2}_{(\l')}}{\MZp^2} \\
     \sum_{\l^0} \e^{\n_1}_{(\l^0)} \e^{* \n_2}_{(\l^0)} &=& - \eta^{\n_1 \n_2} + \frac{k^{\n_1}_{(\l^0)} k^{\n_2}_{(\l^0)}}{\MZO^2} \\
     \sum_{\l_\g} \e^{\m_1}_{(\l^\g)} \e^{* \m_2}_{(\l^\g)} &\to& - \eta^{\m_1\m_2} \label{helicities}
    \eea
where (\ref{helicities}) gives only the relevant part of the sum over
helicities. Other terms are omitted since they give vanishing
contributions to the decay.

The amplitude is given by the sum of the fermionic triangle
$\D_{\r \m \n}^{Z' Z_0 \g}$ plus the proper GCS vertex
    \bea
     A_{\r \m \n}^{Z' Z_0 \g}&=& \D_{\r \m \n}^{Z' Z_0 \g} + (GCS)_{\r \m \n}^{Z' Z_0 \g} \nn\\
             \D_{\r \m \n}^{Z' Z_0 \g}&=&-{1\over4} g_0 g_{Z_0} e \sum_f v_f^{Z'} a_f^{Z_0} q_f \, \G_{\r \m \n}^{VAV}(p,q;m_f)
    \eea
 where $\G_{\r \m \n}^{VAV}(p,q;m_f)$ is given by (\ref{VAVtrian}).
 It is convenient to express the triangle amplitude by using the Rosenberg parametrization~\cite{Rosenberg:1962pp}
    \bea
     \D_{\r \m \n}^{Z' Z_0 \g}&=&
        -{1\over4\pi^2} g_0 g_{Z_0} e \Big(A_1 \e[p,\m,\n,\r] +
         A_2\e[q,\m,\n,\r]+
          A_3 \e[p,q,\m,\r]{p}_{\n} \nn\\
        &&+  A_4 \e[p,q,\m,\r]{q}_{\n} +  A_5 \e[p,q,\n,\r]p_\m +
         A_6\e[p,q,\n,\r]q_\m \Big)
      \eea
where
\be
 A_i=\sum_f v_f^{Z'} a_f^{Z_0} q_f I_i  \qquad \text{for } i=3, \dots , 6
\ee
$I_3,~I_4,~I_5$ and $I_6$ are finite integrals (their explicit forms are given in (\ref{I's}))
and $\e[p,q,\r,\s]$ is defined after (\ref{Ros}). $A_1$ and $A_2$
are naively divergent by power counting and so they must be regularized. We compute them by using the Ward
identities. In this way it is possible to express $A_1$ and $A_2$ in terms of the finite integrals $I_3,~I_4,~I_5$
and $I_6$. The GCS term has the following tensorial structure
\be
d^{Z' Z_0 \g}\Big(\e[p,\m,\n,\r]-\e[q,\m,\n,\r] \Big)
\ee
so it can be absorbed by shifting the first two coefficients of
the Rosenberg parametrization for the triangle. The resulting
amplitude can be written as
    \bea
     \D_{\r \m \n}^{Z' Z_0 \g}&=&
        -{1\over4\pi^2} g_0 g_{Z_0} e \Big(\tilde A_1 \e[p,\m,\n,\r] +
         \tilde A_2\e[q,\m,\n,\r]+
          A_3 \e[p,q,\m,\r]{p}_{\n} \nn\\
        &&+  A_4 \e[p,q,\m,\r]{q}_{\n} +  A_5 \e[p,q,\n,\r]p_\m +
         A_6\e[p,q,\n,\r]q_\m \Big) \label{ampZ'Zg}
      \eea
The Ward identities (\ref{GoldWI}) on the amplitude now read
     \bea
     (p+q)^\r A_{\r \m \n}^{Z' Z_0 \g}+i M_{Z'} (GS)^{Z_0 \g}_{\m \n}&=&0\\
      p^\m A_{\r \m \n}^{Z' Z_0 \g}+i M_{Z_0} (NG)^{Z' \g}_{\r \n}&=&0\\
      q^\n  A_{\r \m \n}^{Z' Z_0 \g}&=& 0
    \eea
where $M_{Z'}= 4b_3 g_0$ and $M_{Z_0}$ are the $Z'$ and $Z_0$ masses respectively. After some manipulations we
obtain
\bea
 (p+q)^\r A_{\r \m \n}^{Z' Z_0 \g}&=&
  {1\over4\pi^2} g_0 g_{Z_0} e {1\over2}\sum_f v_f^{Z'} a_f^{Z_0} q_f ~\e[p,q,\m,\n]\\
 p^\m A_{\r \m \n}^{Z' Z_0 \g}&=&
   -{1\over4\pi^2} g_0 g_{Z_0} e \sum_f v_f^{Z'} a_f^{Z_0} q_f m_f^2 ~I_0 ~\e[q,p,\n,\r]\\
 q^\n A_{\r \m \n}^{Z' Z_0 \g}&=& 0
\eea
and inserting (\ref{ampZ'Zg}) into the above identities we get
    \bea
      \tilde A_1 &=&  \(q^2 A_4 + p \cdot q A_3 \) \nn\\
       \tilde A_2 &=& \(p^2 A_5 + p \cdot q A_6 + (NG)^{Z' \g} \)\label{Z'ZgA1}
    \eea
with
\be
 (NG)^{Z' \g}=\sum_f v_f^{Z'} a_f^{Z_0} q_f \ m_f^2 I_0
\ee where $I_0$ is the integral given in (\ref{I_0integral}). Substituting $\tilde A_1, ~ \tilde A_2$ from (\ref{Z'ZgA1}) into the amplitude
(\ref{ampZ'Zg}) and performing all the contractions we finally obtain
    \bea
     |A^{Z' Z_0 \g}|^2 &=& g_0^2 g_{Z_0}^2 e^2 \frac{
     \left(M_{Z'}^2-M_{Z_0}^2\right)^2 \left(M_{Z'}^2+M_{Z_0}^2\right)}{96 M_{Z_0}^2 M_{Z'}^2 \pi^4} \times\nn\\
    &&
    \[\sum_f   v_f^{Z'} a_f^{Z_0} q_f  \Big(  (I_3+I_5) M_{Z_0}^2 + m_f^2 ~I_0  \Big)\]^2 \label{Z'Zg}
   \eea

   \subsection{$ Z' \to Z_0 \ Z_0 $}

The computations are similar to the previous case so we point out
only the differences with the other decay. Mutatis mutandis, the
decay rate for the process is given in (\ref{decay1}) with
 \be
     p_F = \frac{\MZp}{2} \sqrt{ 1 - \frac{4\MZO^2}{\MZp^2} }
    \label{decay12}\ee
    The square of the total scalar amplitude is given by
    \be
     |A_\text{TOT}|^2 =  \frac{1}{3} \sum_{\l'} \e^{\r_1}_{(\l')} \e^{* \r_2}_{(\l')} \
                       \sum_{\l^0} \e^{\n_1}_{(\l^0)} \e^{* \n_2}_{(\l^0)} \
                      \sum_{\l^0} \e^{\m_1}_{(\l^0)} \e^{* \m_2}_{(\l^0)} \
                      A_{\r_1 \m_1 \n_1}^{Z' Z_0 Z_0} A_{\r_2 \m_2 \n_2}^{* \, Z' Z_0 Z_0}
    \ee
where the amplitude $A_{\r \m \n}$ is always the sum of the
fermionic triangle and the (GCS) term. The contribution to the
fermionic triangle is%
\bea
 \D_{\r \m \n}^{Z' Z_0 Z_0}&=&-{1\over8} g_0 g_{Z_0}^2 \Bigg[\sum_f \(
     v_f^{Z'} a_f^{Z_0} v_f^{Z_0} \, \G_{\r \m \n}^{VAV} +
     v_f^{Z'} v_f^{Z_0} a_f^{Z_0} \, \G_{\r \m \n}^{VVA} \)+ \nn\\
    &&\qquad \quad +\sum_n \( a_n^{Z'} v_n^{Z_0} v_n^{Z_0} \, \G_{\r \m \n}^{AVV}+
     a_n^{Z'} a_n^{Z_0} a_n^{Z_0} \, \G_{\r \m \n}^{AAA} \) \Bigg]
\eea
where $n$ runs over all the neutrinos while the $\G_{\r \m \n}$'s are given by (\ref{AVVtrian}), (\ref{AAAtrian}), (\ref{VAVtrian}), (\ref{VVAtrian}). Using the fact that for
the three neutrino families
we have $v_n^{Z'}=a_n^{Z'}$ and $v_n^{Z_0}=a_n^{Z_0}$ we write the total amplitude (the sum of the triangles plus GCS terms) as
\bea
     A_{\r \m \n}^{Z' Z_0 Z_0}&=&
        -{1\over8\pi^2} g_0 g_{Z_0}^2  \Big(\tilde A_1 \e[p,\m,\n,\r] +
         \tilde A_2 \e[q,\m,\n,\r]+
          A_3 \e[p,q,\m,\r]{p}^{\n} \nn\\
        &&+  A_4 \e[p,q,\m,\r]{q}^{\n} +  A_5 \e[p,q,\n,\r]p^\m +
         A_6\e[p,q,\n,\r]q^\m \Big)
\eea
with
\be
 A_i=2 \sum_f \tilde v_f^{Z'} a_f^{Z_0} v_f^{Z_0} I_i  \qquad \text{for } i=3, \dots , 6
\ee
where $\tilde v_n^{Z'}=2 v_n^{Z'}$ for neutrinos and
$\tilde v_f^{Z'}=v_f^{Z'}$ for the other fermions. The Ward identities
now read
\bea
     (p+q)^\r A_{\r \m \n}^{Z' Z_0 Z_0} +i M_{Z'} (GS)^{Z_0 Z_0}_{\m \n}&=&0\\
      p^\m A_{\r \m \n}^{Z' Z_0 Z_0} +i M_{Z_0} (NG)^{Z' Z_0}_{\r \n}&=&0\\
      q^\n A_{\r \m \n}^{Z' Z_0 Z_0} +i M_{Z_0} (NG)^{Z_0 Z'}_{\m \r}&=&0
    \eea
leading to
\bea
 (p+q)^\r A_{\r \m \n}^{Z' Z_0 Z_0}&=&
  {1\over8\pi^2} g_0 g_{Z_0}^2 \sum_f \tilde v_f^{Z'} a_f^{Z_0} v_f^{Z_0} \e[p,q,\m,\n]\\
 p^\m A_{\r \m \n}^{Z' Z_0 Z_0}&=&
   -{1\over8\pi^2} g_0 g_{Z_0}^2 \sum_f \tilde v_f^{Z'} a_f^{Z_0} v_f^{Z_0} m_f^2 I_0 \e[q,p,\n,\r]\\
 q^\n A_{\r \m \n}^{Z' Z_0 Z_0}&=&
   -{1\over8\pi^2} g_0 g_{Z_0}^2 \sum_f \tilde v_f^{Z'} a_f^{Z_0} v_f^{Z_0} m_f^2 I_0 \e[q,p,\r,\m]
\eea
From these equations we find the following values for $\tilde A_1$ and $\tilde A_2$
\bea \tilde A_1 &=&
   \(q^2 A_4 + p \cdot q A_3 - (NG)^{Z' Z_0} \) \\
   \tilde A_2 &=& \(p^2 A_5 + p \cdot q A_6 + (NG)^{Z' Z_0}\)
\eea
with
\be
 (NG)^{Z' Z_0}=\sum_f \tilde v_f^{Z'} a_f^{Z_0} v_f^{Z_0} \ m_f^2 I_0
\ee
where $I_0$ is the integral given in (\ref{I_0integral}). Substituting back into the amplitude and performing all the contractions we finally
obtain
\bea
 |A^{Z' Z_0 Z_0}|^2 &=&
 g_0^2 g_{Z_0}^4 \frac{ \left(M_{Z'}^2-4 M_{Z_0}^2\right)^2}{192 M_{Z_0}^2\pi^4}
 \[ \sum_f \tilde v_f^{Z'} a_f^{Z_0} v_f^{Z_0}
\bigg( 2(I_3+I_5) M_{Z_0}^2+ m_f^2 I_0 \bigg)\]^2\label{ampzprimozozo}
\eea

\subsection{Numerical Results}\label{Plots}

\begin{figure}[p]
      \centering
      \includegraphics[scale=0.38]{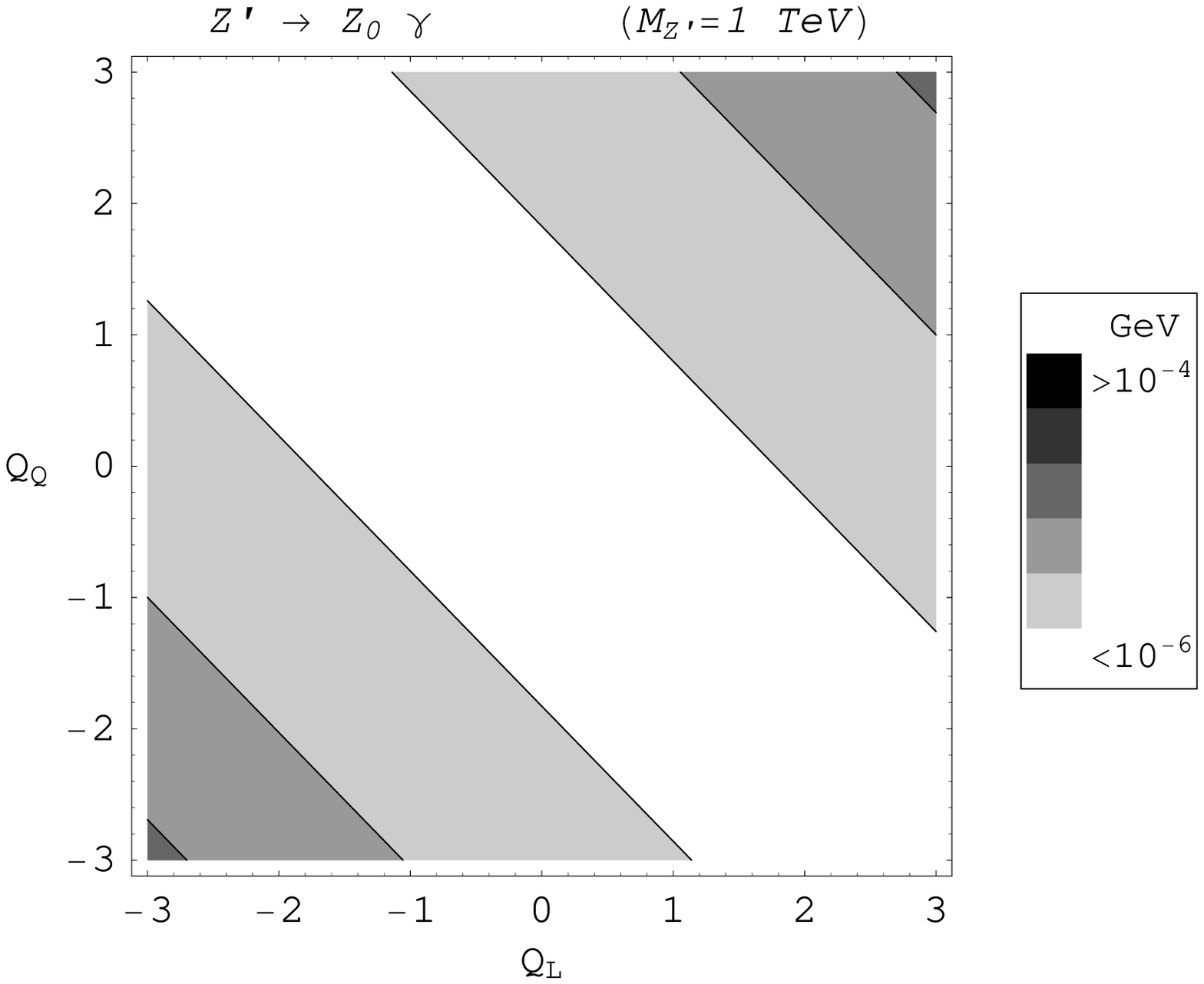}
      ~~~~~~
      \includegraphics[scale=0.38]{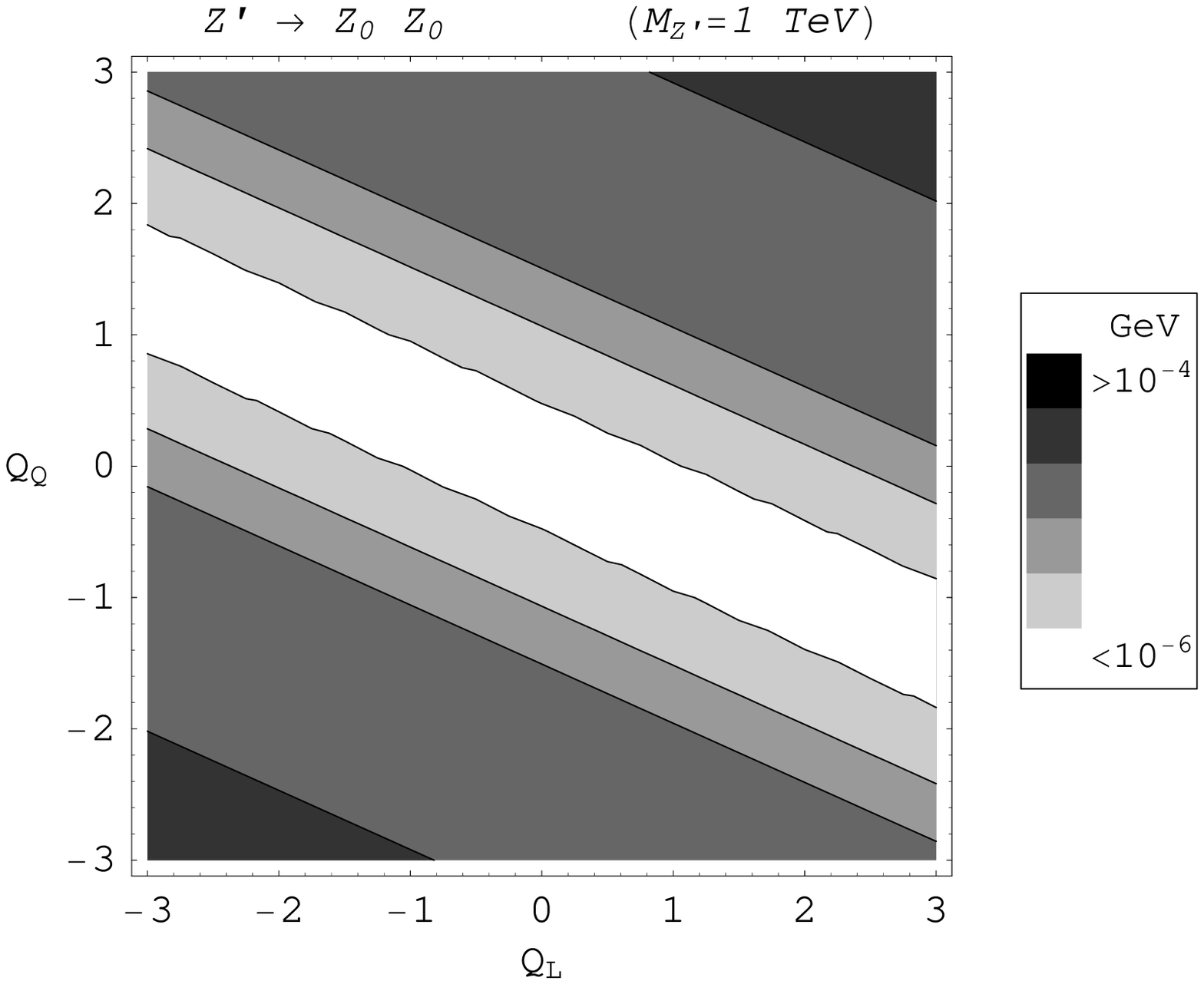}
      \caption{$\MZp=1$ TeV.} \label{Contourplot 1 TeV}
     \end{figure}

\begin{figure}[hp]
      \centering
      \includegraphics[scale=0.38]{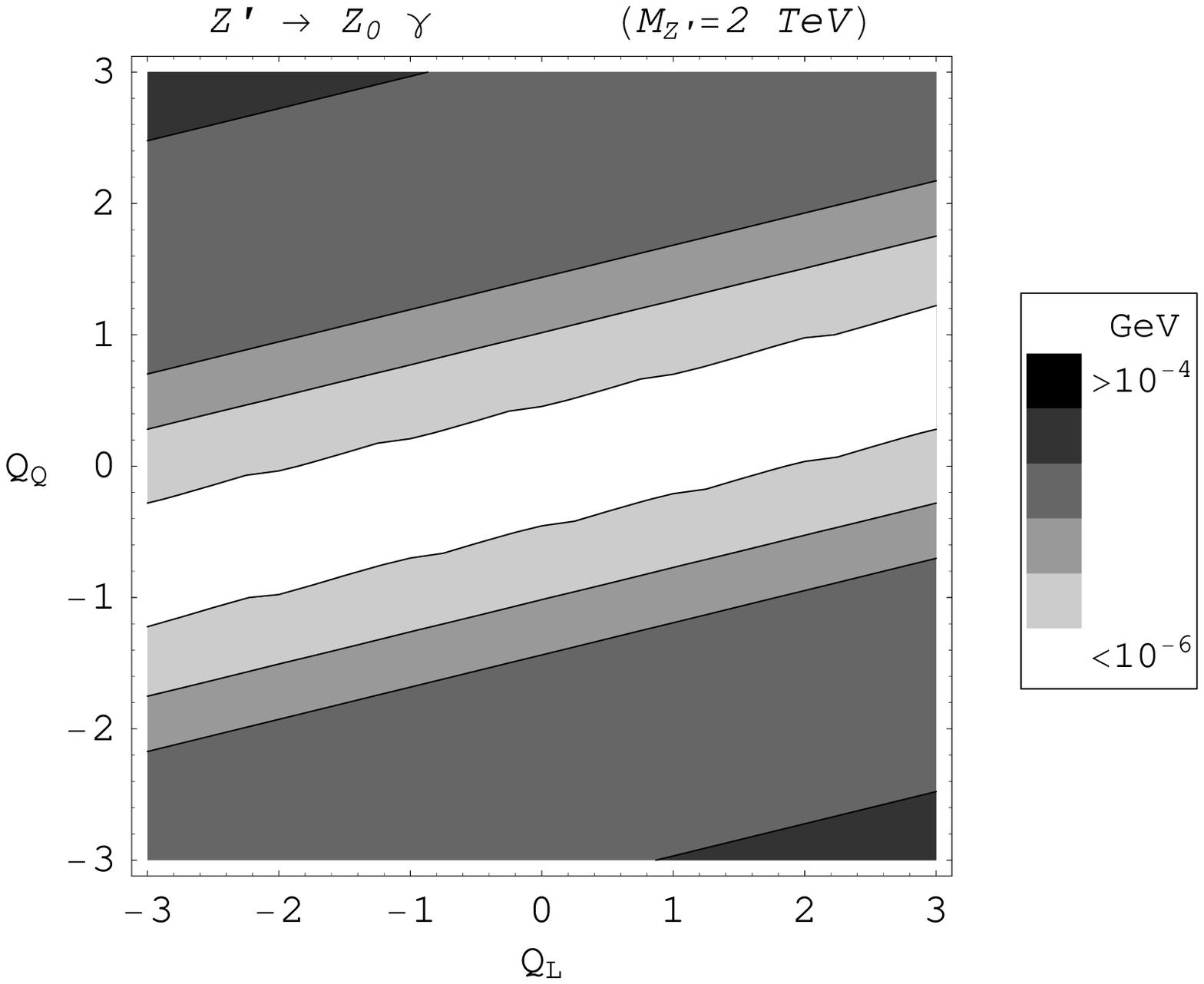}
      ~~~~~~
      \includegraphics[scale=0.38]{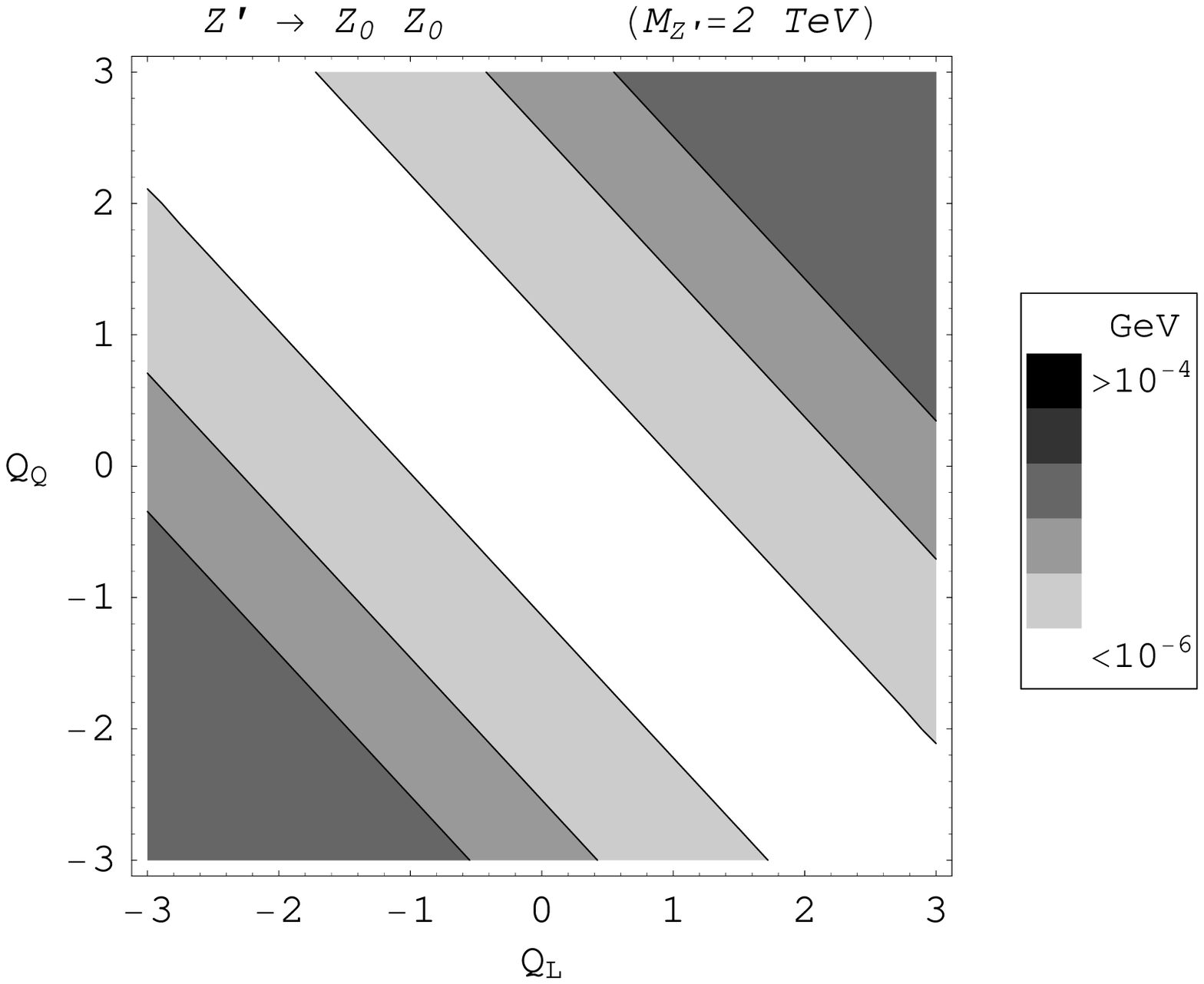}
      \caption{$\MZp=2$ TeV.} \label{Contourplot 2 TeV}
     \end{figure}

\begin{figure}[hp]
      \centering
      \includegraphics[scale=0.38]{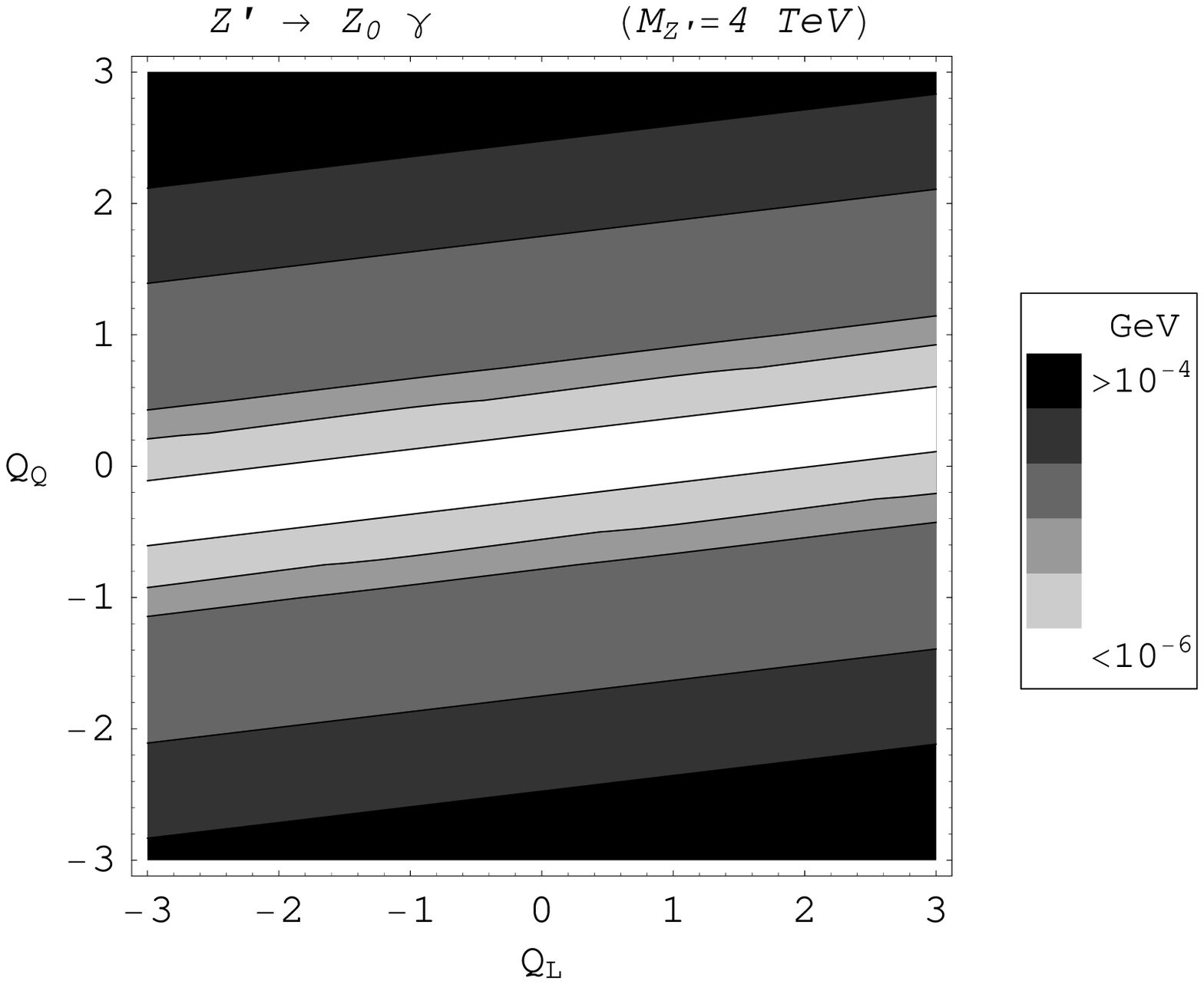}
      ~~~~~~
      \includegraphics[scale=0.38]{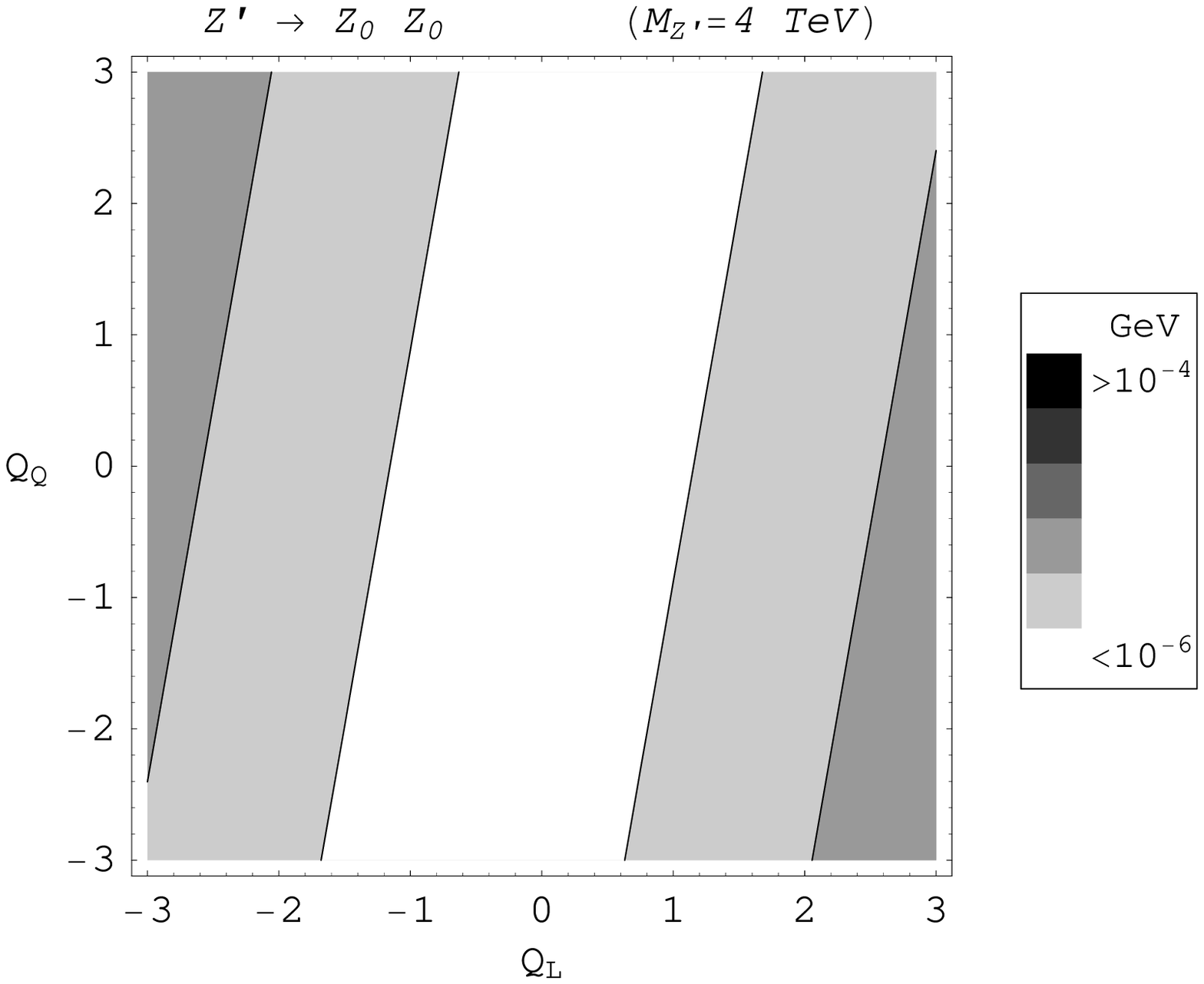}
      \caption{$\MZp=4$ TeV.} \label{Contourplot 4 TeV}
     \end{figure}

In this Section we show some numerical computations for the two decay rates $\Gamma
(Z'\to Z_0 \gamma)$ and $\Gamma (Z'\to Z_0 Z_0)$. They depend on the free parameters
of the model, i.e. the charges $Q_Q$, $Q_L$ and the mass of the $Z'$. We assume that
$Q_{H_u}=0$ and we choose $g_0=0.1$.
We show our results in Fig. \ref{Contourplot 1 TeV}-\ref{Contourplot 4 TeV}
in the form of contour plots in the plane $Q_Q, Q_L$ for $M_{Z'}=1,2$ and $4$ TeV.
Our choices for $g_0$, $Q_Q$, $Q_L$ and $\MZp$ are in agreement with the current experimental bounds \cite{CDF}.

The darker shaded regions correspond to larger decay rates. The white region
corresponds to the value $10^{-6}$ GeV that can be considered as a rough lower
limit for the detection of the corresponding process.
It is worth noting that increasing $M_{Z'}$ the mean value of the decay rate of $Z'\to Z_0 \gamma$
grows while the one of $Z'\to Z_0 Z_0$ decreases.
We would also like to mention that increasing $M_{Z'}$ the iso-decay rate contours in the plot rotate clockwise getting  more and more parallel to the $Q_L$-axis. This effect is due to
the fact that the contribution of the triangle diagram with the top quark
circulating inside the loop becomes the
dominant contribution for high $M_{Z'}$.
In this case the decays strongly depend on the top quark charge
$Q_Q$ while the lepton charges $Q_L$ become irrelevant.
Finally, we find that the region that gives the largest values (of order of $10^{-4}$ GeV) of the decay $Z' \to Z_0 \, \gamma$
is for $M_{Z'}\sim 4$ TeV and for $Q_Q\sim 3$, $Q_L\sim -2$.

\begin{figure}[t]
      \centering
      \includegraphics[scale=0.38]{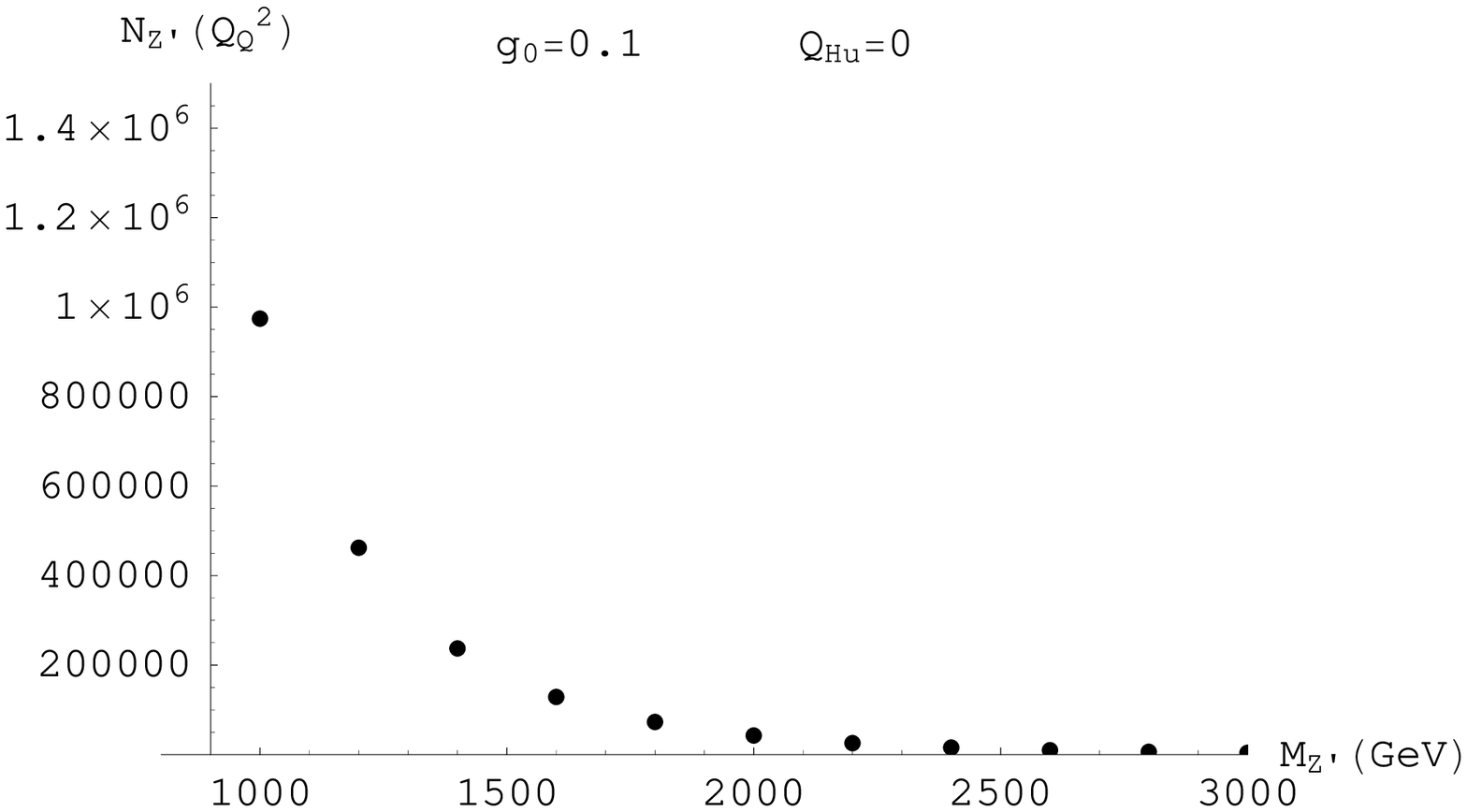}
      \caption{Number of $Z'$ produced at LHC in 1 year for $\L=10^{34} cm^{-2} s^{-1}$ and $\sqrt s = 14$ TeV,
               in units of $Q_Q^2$, in function of the mass of the $Z'$.} \label{NZp}
     \end{figure}

\begin{figure}[h]
      \centering
      \includegraphics[scale=0.38]{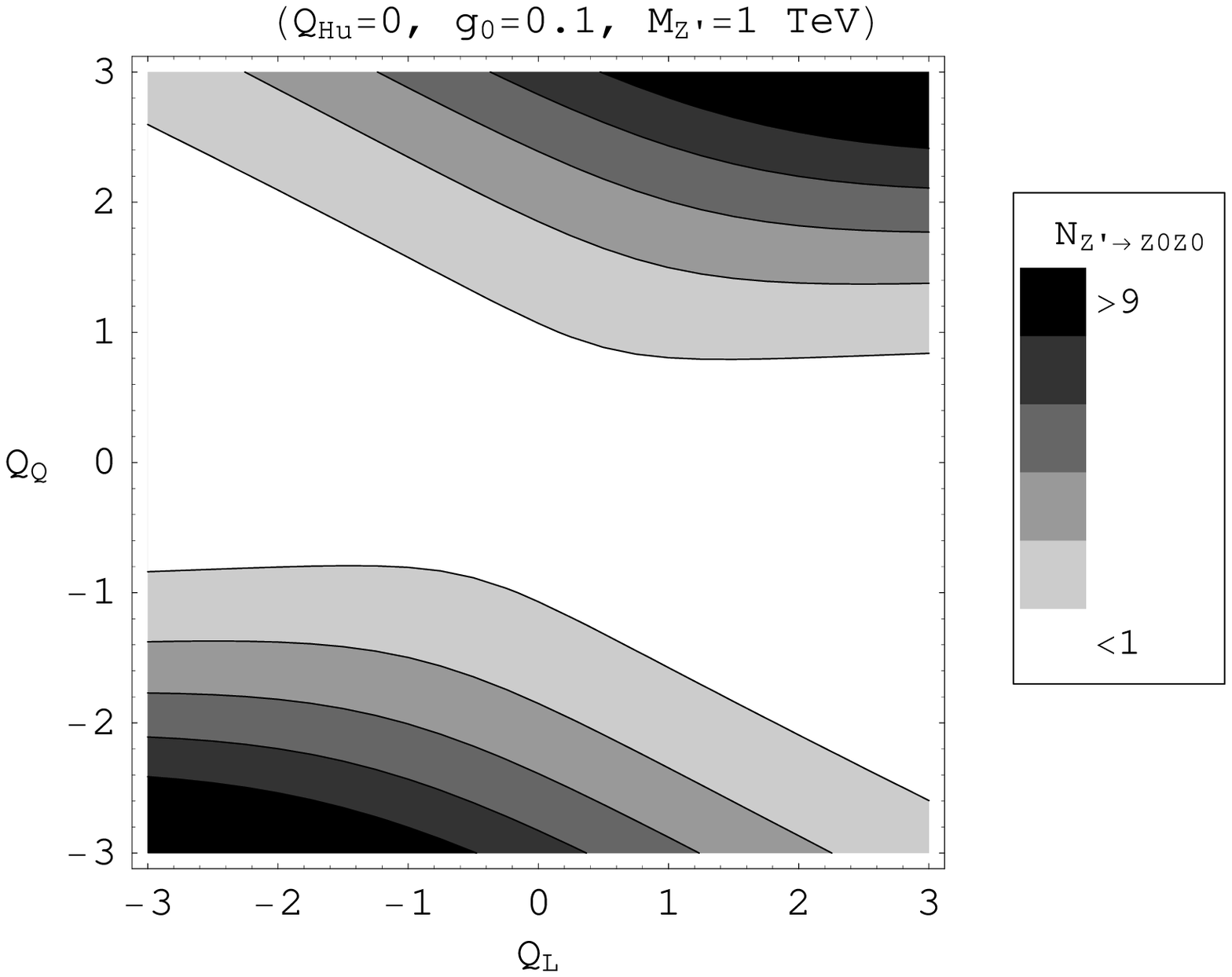}
      \caption{Number of $Z'\to Z_0 Z_0$ at LHC in 1 year for $\L=10^{34} cm^{-2} s^{-1}$, $\sqrt s = 14$ TeV and
               $\MZp=1$ TeV.} \label{NZpZZ}
     \end{figure}

To estimate the number of the anomalous decays that can be observed at LHC we shall use the
narrow width approximation,
\be
N_{Z' \to \text{particles}}=
N_{Z'} \ \text{BR} (Z' \to \text{particles})
\ee
where $N_{Z'}=\s_{Z'} \, \L \, t \ $ is the total number of $Z'$,
$\text{BR} (Z' \to \text{particles})$ is the branching ratio,
$\L=10^{34} {\rm \,\,cm^{-2} s^{-1}}$ the luminosity and $t=$1 year.
Finally $\s_{Z'}$ is the $Z'$ production cross section \cite{Langacker:2008yv}
\be
\frac{d\s_{Z'}}{dy}= \frac{4\pi^2 x_1 x_2}{3M_{Z'}^3} \sum_{i}
\big[f_{q_i}(x_1)f_{\bar q_i}(x_2)
                      +f_{\bar q_i}(x_1)f_{q_i}(x_2)\big] \Gamma (Z' \to q_i \bar q_i),
\label{Zpproduction}
\ee
where $f_{q_i,\bar q_i}$ are the quark $q_i$ (or
antiquark $\bar q_i$) structure functions   in the
proton, and the momentum fractions are
\be
 x_{1,2}=(M_{Z'}/\sqrt{s}) e^{\pm y}.
\label{xval} \ee
To estimate a rough  upper bound for the anomalous BR we
assume that the sfermions will have an universal mass of about 500 GeV. We integrate numerically the PDFs
using a Mathematica package \cite{PDF}. In Fig. \ref{NZp} we show the result for $N_{Z'}$ at $\sqrt s = 14$ TeV.
We can see that the number of the $Z'$ produced falls off
exponentially with $M_{Z'}$, so we shall focus on the case $M_{Z'} \sim 1$
TeV and the most favorite decay $Z'\to Z_0 Z_0$. In Fig. \ref{NZpZZ}, we estimate the number of decays for 1 year
of integrated luminosity which turns out to be
$N_{Z'\to Z_0 Z_0} \sim 10$ for large values of the charges $\QL$ and $\QQ$.
We will present a more detailed analysis in a forthcoming paper \cite{inprogress}.
\newpage
\vskip 2cm
\begin{flushleft}
{\large \bf Acknowledgments}
\end{flushleft}

\noindent It is a pleasure to thank Massimo Bianchi, Claudio
Corian\'o, Anna Di Ciaccio,  Gennaro Corcella, Giorgio Chiarelli, Marco Guzzi, Marco
Zagermann.  A special thanks goes to Theodore Tomaras for sharing  with us
his insights on the decay rates we computed. P.A. would
like to thank also Ignatios Antoniadis, Ralph Blumenhagen, Elias
Kiritsis, Oleg Ruchayskiy for useful discussions and CERN and
\'Ecole Polytechnique for hospitality. This work was supported in
part by the CNRS PICS no. 2530 and 3059, INTAS grant 03-516346,
MIUR-COFIN 2003-023852, NATO PST.CLG.978785, the RTN grants
MRTNCT- 2004-503369, EU MRTN-CT-2004-512194, MRTN-CT-2004-005104
and by a European Union Excellence Grant, MEXT-CT-2003-509661.

\vskip 2cm

\appendix

\section{Conventions} \label{appConventions}

We use the space-time metric $\eta_{\m \n} = \text{diag}(+,-,-,-)$ and the spinorial conventions
  \be
   \e_{21}=\e^{12}=1 \qquad \e_{12}=\e^{21}=-1 \qquad \e_{11}=\e^{11}=\e_{22}=\e^{22}=0
  \ee
  \be
   \psi^\a = \e^{\a \b} \psi_\b \qquad \psi_\a = \e_{\a \b} \psi^\b \qquad
   \bar\psi^\ad = \e^{\ad \bd} \bar\psi_\bd \qquad \bar\psi_\ad = \e_{\ad \bd} \bar\psi^\bd
  \ee
  \be
   \psi \chi = \psi^\a \chi_\a   \qquad \bar\psi \bar\chi = \bar\psi_\ad \bar\chi^\ad
  \ee
The Dirac matrices are
\be
\g^\m=\( \begin{array}{cc}   0   &  \s^\m \\
                         \sb^\m  &  0 \end{array} \) \qquad \text{where} \quad
\left\{\begin{array}{rcl} \s^\m&=&(1,-\vec\s)\\
                         \sb^\m&=&(1,\vec\s) \ \end{array} \right.
\ee
and we define
\be
\g_5=\( \begin{array}{cc} -1 &  0 \\
                           ~0 &  1 \end{array} \)
\ee

\section{Total lagrangian} \label{applagrangian}
 The lagrangian of the model contains several
terms
  \be \label{lagrangian}
   \L = \L_Q + \L_L + \L_{gauge} + \L_H + \L_{W} + \L_{axion} + \L_{GCS} + \L_{Soft}
  \ee
where
  \bea
   \L_Q &=& \( Q_i^\dag e^{V^{(3)}} e^{V^{(2)}} e^{{1\over6}V^{(1)}} e^{Q_{Q_i} V^{(0)}}Q_i \right. \nn\\
             &&+\left.(U^c_i)^\dag e^{-V^{(3)}} e^{-{2\over3}V^{(1)}} e^{Q_{U^c_i} V^{(0)}}U^c_i +
               (D^c_i)^\dag e^{-V^{(3)}} e^{{1\over3}V^{(1)}} e^{Q_{D^c_i} V^{(0)}}D^c_i\)_{\thth}
   \eea
   \bea
   \L_L &=& \( L_i^\dag e^{V^{(2)}} e^{-{1\over2}V^{(1)}} e^{Q_{L_i} V^{(0)}}L_i + (E^c_i)^\dag e^{V^{(1)}} e^{Q_{E^c_i} V^{(0)}}E^c_i \)_{\thth}   \eea
   \bea
   \L_H &=& \( H_u^\dag e^{V^{(2)}} e^{{1\over2}V^{(1)}} e^{Q_{H_u} V^{(0)}}H_u + H_d^\dag e^{V^{(2)}} e^{-{1\over2}V^{(1)}} e^{Q_{H_d} V^{(0)}}H_d \)_{\thth}
      \eea
      \bea
   \L_{gauge} &=& \( {1\over8 g_3^2} \Tr\( W^{(3)} W^{(3)} \) + {1\over8 \tilde g_2^2} \Tr\( W^{(2)} W^{(2)} \)  \right. \nn\\
         &&+ \left.{1\over16 \tilde g_1^2} \Tr\( W^{(1)} W^{(1)} \) + {1\over16 (\tilde g_0)^2} \Tr\( W^{(0)} W^{(0)} \)\)_{\th^2} +h.c.
      \eea
      \bea
   \L_{W} &=&\(  y_u^{i j} Q_i U^c_j H_u - y_d^{i j} Q_i D^c_j H_d - y_e^{i j} L_i E^c_j H_d + \m H_u H_d\)_{\th^2} + h.c.  \eea
  \bea
   \L_{axion} &=& {1\over4} \left. \( S + \bar S + 4 b_3 V^{(0)} \)^2 \right|_{\thth} \nn\\
                &&- {1\over2} \left\{ \[\sum_{a=0}^2 b^{(a)}_2 S ~\Tr\( W^{(a)} W^{(a)} \) + b^{(4)}_2 S ~W^{(1)} ~W^{(0)} \]_{\th^2} +h.c. \right\}~
  \eea
  \bea
   \L_{GCS} &=&- d_4     \[ \( V^{(1)} D^\a V^{(0)} - V^{(0)} D^\a V^{(1)}\) W^{(0)}_\a + h.c. \]_{\thth} +\nn\\
             &&+  d_5     \[ \( V^{(1)} D^\a V^{(0)} - V^{(0)} D^\a V^{(1)}\) W^{(1)}_\a + h.c. \]_{\thth} +\nn\\
             &&+  d_6 \Tr \bigg[ \( V^{(2)} D^\a V^{(0)} - V^{(0)} D^\a V^{(2)}\) W^{(2)}_\a +\nn\\
                          &&\qquad \quad+{1\over6} V^{(2)} D^\a V^{(0)} \bar D^2 \(  \[D_\a V^{(2)},V^{(2)}\] \) + h.c. \bigg]_{\thth}
    \eea
    \bea
   \L_{Soft} &=& - {1\over2}  \(\sum_{a=0}^3 M_a \l^{(a)} \l^{(a)} + h.c. \)- {1\over2}  \(M_S \psi_S \psi_S  + h.c. \) \nn\\
               &&   - \( m^2_{Q_{ij}} \Qt_i \Qt_j^* + m^2_{U_{ij}} \Ut_i \Uts_j + m^2_{D_{ij}} \Dt_i \Dts_j  \right. \nn\\
                    &&\left.+m^2_{L_{ij}} \Lt_i \Lt_j^* + m^2_{E_{ij}} \Et_i \Ets_j + m^2_{h_u} |h_u|^2 + m^2_{h_d} |h_d|^2 \) \nn\\
                    &&-\( a_u^{ij} \Qt_i \Ut_j h_u - a_d^{ij} \Qt_i \Dt_j h_d - a_e^{ij} \Lt_i \Et_j h_d + b h_u h_d + h.c. \)\eea
where $\L_Q$, $\L_L$ and $\L_H$ provide the kinetic terms and the
gauge interactions of the matter particles such as (s)quarks,
(s)leptons, Higgs(ino)s; $\L_{gauge}$ contains the kinetic terms
for the gauge supermultiplet; $\L_W$ is the usual MSSM
superpotential; $\L_{axion}$ provides the kinetic term of the
St\"uckelberg multiplet and its Green-Schwarz interactions used
in the anomaly cancellation procedure; $\L_{GCS}$ contains the
Generalized Chern Simons interactions giving trilinear gauge boson
couplings needed to complete the anomaly cancellation procedure;
finally, $\L_{Soft}$ contains the usual soft breaking
terms of the MSSM as well as the new terms for the primeino and the
axino.

Notice that in order to include the coupling constants in the gauge
interactions we need to redefine them as shown in equation
(\ref{couplingconst}) and to substitute $V \to 2 g V$.

 \section{Amplitudes, Ward identities and Anomalies} \label{appendixanomalies}

\subsection{Fermionic loop diagram} \label{fermionic triangle}

     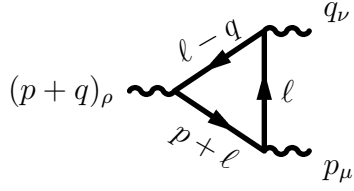
\begin{figure}[tb]
\vskip 1.5cm
      \centering
      \raisebox{-4.3ex}[0cm][0cm]{\unitlength=0.4mm
      \begin{fmffile}{loop3a}
      \begin{fmfgraph*}(60,40)
       \fmfpen{thick} \fmfleft{ii0,ii1,ii2} \fmfstraight \fmffreeze \fmftop{ii2,t1,t2,t3,oo2}
       \fmfbottom{ii0,b1,b2,b3,oo1} \fmf{phantom}{ii2,t1,t2} \fmf{phantom}{ii0,b1,b2} \fmf{phantom}{t1,v1,b1}
       \fmf{phantom}{t2,b2} \fmf{phantom}{t3,b3}
       \fmffreeze
       \fmf{photon}{ii1,v1}
       \fmf{fermion,label.side=right,label=\begin{rotate}{30}$\! \! \! \! \! \! \ell-q$\end{rotate}}{t3,v1}
       \fmf{fermion,label=$\ell$}{b3,t3}
       \fmf{fermion,label.dist=0.5cm,label=\begin{rotate}{-30} $\! \! \! \! \! \! p+\ell$\end{rotate}}{v1,b3}
       \fmf{photon}{b3,oo1} \fmf{photon}{t3,oo2}
       \fmflabel{$(p+q)_\r$}{ii1} \fmflabel{$p_\m$}{oo1} \fmflabel{$q_\n$}{oo2}
      \end{fmfgraph*}
      \end{fmffile}}
    \vskip 1.5cm
    \caption{The anomalous triangle diagram.}\label{TriangleDiagram}\end{figure}

In this Subsection we give some general properties of the
fermionic triangle diagram of Fig.~\ref{TriangleDiagram}. Consider
a case in which only a single fermion circulates in the loop and
each coupling is either axial (A) or vectorial (V) with charge
equal to minus one. The fermionic triangles containing  an odd
number of axial couplings, denoted by AVV , VAV, VVA and AAA are
\bea \G_{\r \m \n}^{AVV}(p,q;m_f)&=&\int \frac{d^{4} \ell}{(2
\pi)^{4}} \, Tr\(  \g_5  \g_\r  \frac{1}{\slashed \ell - \slashed
q - m_f} \g_\n \frac{1}{\slashed \ell -m_f} \g_\m
\frac{1}{\slashed \ell + \slashed p - m_f } \)+    \nn\\
&&+ (p \leftrightarrow q, \m \leftrightarrow \n) \label{AVVtrian}\\
  \G_{\r \m \n}^{VAV}(p,q;m_f)&=&\int \frac{d^{4} \ell}{(2 \pi)^{4}}
  \, Tr\(  \g_\r  \frac{1}{\slashed \ell - \slashed q - m_f}
  \g_\n \frac{1}{\slashed \ell -m_f} \g_5 \g_\m
  \frac{1}{\slashed \ell + \slashed p - m_f } \)+    \nn\\
  &&+ (q \leftrightarrow -(p+q), \n \leftrightarrow \r) \label{VAVtrian}\\
  \G_{\r \m \n}^{VVA}(p,q;m_f)&=&\int \frac{d^{4} \ell}{(2 \pi)^{4}}
  \, Tr\(   \g_\r  \frac{1}{\slashed \ell - \slashed q - m_f}
  \g_5 \g_\n  \frac{1}{\slashed \ell -m_f}  \g_\m
  \frac{1}{\slashed \ell + \slashed p - m_f } \)+\nn\\
  &&+ (p \leftrightarrow -(p+q), \m \leftrightarrow \r)  \label{VVAtrian}\\
\G_{\r \m \n}^{AAA}(p,q;m_f)&=&\int \frac{d^{4} \ell}{(2 \pi)^{4}}
\, Tr\(  \g_5  \g_\r  \frac{1}{\slashed \ell - \slashed q - m_f}
\g_5  \g_\n \frac{1}{\slashed \ell -m_f} \g_5 \g_\m
\frac{1}{\slashed \ell + \slashed p - m_f } \)+\nn\\
&&+ (p \leftrightarrow q, \m \leftrightarrow \n)  \label{AAAtrian}
\eea
These integrals are superficially divergent (by power counting)
and thus there is an ambiguity in their definition. The internal
momentum $\ell$ can, in fact, be arbitrarily
shifted~(see Section 6.2 of \cite{Cheng:1985bj}) \be
 \ell_\s \to \ell_\s+\a\, p_\s+(\a-\b)q_\s
\ee
leading to
\be \G_{\r \m \n}^{AVV}(p,q,\b;m_f)=\G_{\r \m
\n}^{AVV}(p,q;m_f)-\frac{\b}{8\pi^2}\epsilon_{\r\m\n\s}(p-q)^\s
\label{Gshift}
\ee
The amplitudes
(\ref{AVVtrian}),(\ref{VAVtrian}),(\ref{VVAtrian}) and
(\ref{AAAtrian}) can be written using the the Rosenberg
parametrization \cite{Rosenberg:1962pp} as
\bea
&&\G_{\r \m \n}  (p,q;m_f)=~~~\nn\\
&&~~~{1\over{ \pi^2}} \Big(
I_1(p,q;m_f) \,\e[p,\m,\n,\r] + I_2(p,q;m_f) \, \e[q,\m,\n,\r]+  I_3(p,q;m_f)  \, \e[p,q,\m,\r]{p}_{\n} \nn\\
&&~~~~+  I_4(p,q;m_f) \, \e[p,q,\m,\r]{q}_{\n} +  I_5(p,q;m_f) \, \e[p,q,\n,\r]p_\m +
I_6(p,q;m_f) \, \e[p,q,\n,\r]q_\m \Big)  \nn\\
\label{Ros}
\eea
with $\e[p,q,\r,\s]=\epsilon_{\m\n\r\s} p^\m q^\n$ and where
\bea I_3(p,q;m_f) &=& -\int_0^1 dx
\int_0^{1-x} dy \frac{x y}{y (1-y) p^2 +
  x(1-x) q^2 + 2 x y \,p\cdot q - m_f^2}  \nonumber \\
I_4(p,q;m_f) &=& \int_0^1 dx \int_0^{1-x} dy \frac{x (x-1)}{y (1-y) p^2 +
  x(1-x) q^2 + 2 x y \,p\cdot q - m_f^2}  \nonumber \\
I_5(p,q;m_f) &=& -I_4(q,p;m_f) \nonumber \\
I_6(p,q;m_f) &=& -I_3(p,q;m_f)  \label{I's}
\eea
In terms of the Rosenberg parametrization  the $\b$ dependence of (\ref{Gshift}) is
contained only in $I_1$ and $I_2$ ( which are superficially
divergent). However, using the Ward identities,
\bea
(p+q)^\r \G_{\r \m \n}^{AVV}(p,q,\b;m_f) &=& {1\over{\pi^2}} \[ {\b\over4} + m_f^2 I_0 (p,q;m_f)\] \e[p,q,\m,\n] \nn \\
p^\m \G_{\r \m \n}^{AVV}(p,q,\b;m_f)     &=& -{2+\b\over8\pi^2}\,  \e[q,p,\n,\r] \nn \\
q^\n \G_{\r \m \n}^{AVV}(p,q,\b;m_f)     &=& -{2+\b\over8\pi^2}\,  \e[q,p,\r,\m]
\eea
where $I_0$ is defined in (\ref{I_0integral}),
 it is possible to
show that they can be expressed in terms of $I_3 \dots I_6$ as
\bea
I_1^{AVV}(p,q,\b;m_f) &=& p \cdot q \, I_3(p,q) + q^2 \, I_4(p,q) +\frac{2+\b}{8}\nn\\
I_2^{AVV}(p,q,\b;m_f) &=& -I_1^{AVV}(q,p,\b;m_f)
\eea
From now on we omit the explicit $\b$ dependence to get more compact formulae.

\subsection{Anomaly distribution and cancellation.}\label{GCSabsorption}
In this Subsection we show that the sum of the triangle amplitude and of the GCS vertex are independent of $\beta$.
Since the anomaly is independent of the fermion masses we discuss only the unbroken phase, i.e. $m_f=0$. We consider the anomaly between
$V^{(0)}$ and two $V^{(1)}$.
The total fermionic triangle (the sum of AAA+AVV+VAV+VVA triangles) can be written as
     \be
      \D_{\r \m \n}^{011}(p,q;0) = -{\cA^{(1)}\over16}   \G_{\r \m \n}  (p,q;0) \label{DeltaAYY}
     \ee
where $\cA^{(1)}$ is the anomaly (\ref{Triangles2}) and $\G_{\r \m \n} $ is defined in (\ref{Ros}).
The Ward identities for the fermionic triangle are
 \bea
      (p+q)^\r  \D_{\r \m \n}^{011} &=& - \b \, \frac{ \cA^{(1)} }{64 \pi^2} \e_{\m \n \a \b} p^\a q^\b \nn\\
       p^\m  \D_{\r \m \n}^{011} &=& \( 2+\b \)   \frac{ \cA^{(1)} }{128\pi^2}  \e_{\n \r \a \b} q^\a p^\b \nn\\
       q^\n  \D_{\r \m \n}^{011} &=& \( 2+\b \)  \frac{ \cA^{(1)} }{128\pi^2}  \e_{\r \m \a \b} q^\a p^\b
 \eea
For instance, $\b=-2/3$ corresponds to a symmetric distribution of the anomaly.
The gauge invariance of the theory is restored using (see Section (\ref{AnomalyCancellationSymmPhase}))
     \bea
      (p+q)^\r \Big(\D_{\r \m \n}^{011}(p,q;0) +(GCS)^{011}_{\r\m\n}\Big)+2i b_3 (GS)^{11}_{\m\n}&=&0\nn \\
      p^\m  \Big(\D_{\r \m \n}^{011}(p,q;0) +(GCS)^{011}_{\r\m\n}\Big) &=&0\nn \\
      q^\n  \Big(\D_{\r \m \n}^{011}(p,q;0) +(GCS)^{011}_{\r\m\n}\Big) &=&0
     \eea
The last two identities imply
\be
 \( 2+\b \)   \frac{ \cA^{(1)} }{128\pi^2} - 2 d_5 = 0 \qquad \Rightarrow \qquad d_5 = { 2+\b \over 2}   \frac{ \cA^{(1)} }{128\pi^2}
\ee
and the first identity becomes
\be
 - \b \, \frac{ \cA^{(1)} }{64 \pi^2} + 4 \ { 2+\b \over 2}   \frac{ \cA^{(1)} }{128\pi^2} + 4 b^{(1)}_2 b_3 =0
\qquad \Rightarrow \qquad b^{(1)}_2 b_3 =- \frac{\cA^{(1)}}{128 \pi^2}
\ee
It is then clear that different choices in the anomaly distribution affect only the GCS coefficient $d_5$ while the GS coefficient $b^{(1)}_2$
remains the same.
This means that removing the St\"uckelberg coupling by gauge fixing and computing the physical
amplitude $\D+GCS$, we get the same result and the same Ward identity.
Consider the amplitude
  \bea
   A^{011}_{\r \m \n} =\D^{011}_{\r \m \n} + (GCS)^{011}_{\r \m \n}
                =\D^{011}_{\r \m \n} + 2 d_5 \e_{\r \n \m \a} (p-q)^\a
  \eea
The GCS terms can be reabsorbed by the following redefinitions
     \bea
      \( {\cA^{(1)}\over{16 \pi^2}}  \) \tilde I_1(p,q) &=& \( {\cA^{(1)}\over{16 \pi^2}}  \) I_1(p,q)-2 d_5\\
      \( {\cA^{(1)}\over{16 \pi^2}} \) \tilde I_2(p,q) &=& \( {\cA^{(1)}\over{16 \pi^2}}  \) I_2(p,q)+2 d_5    \eea
Imposing the $p^\m$ and $q^\n$ identities (\ref{WardIdAYY}) we get
     \bea
      \tilde I_1(p,q) &=& p \cdot q I_3(p,q) + q^2 I_4(p,q) \nn\\
      \tilde I_2(p,q) &=& -\tilde I_1(q,p) \label{I1tilde}
     \eea
that relate $\tilde I_1$ and $\tilde I_2$ to the other $I_i$'s. We can define
     \bea
     \tilde \G_{\r \m \n}  &=&
      {1\over{ \pi^2}} \Big( \tilde I_1 \e[p,\m,\n,\r] +\tilde I_2\e[q,\m,\n,\r]+ I_3 \e[p,q,\m,\r]{p}^{\n} \nn\\
        &&+  I_4 \e[p,q,\m,\r]{q}^{\n} +  I_5 \e[p,q,\n,\r]p^\m +
         I_6(\e[p,q,\n,\r]q^\m \Big)
     \eea
so that the amplitude is
     \be
      A^{011}_{\r \m \n} =\D^{011}_{\r \m \n} + (GCS)^{011}_{\r \m \n}
                       =-{\cA^{(1)} \over 16 } \tilde \G_{\r \m \n}
     \ee
and obeys the following Ward identities
     \bea
      (p+q)^\r A^{011}_{\r \m \n} &=& \frac{\cA^{(1)}}{32 \pi^2} \e_{\m \n \a \b} p^\a q^\b=-2 i b_3 (GS)^{11}_{\m \n}\nn \\
      p^\m A^{011}_{\r \m \n} &=&0\nn \\
      q^\n A^{011}_{\r \m \n} &=&0
     \eea
This result does not depend on the scheme of the anomaly distribution.

\subsection{Treatment of non anomalous diagrams\label{notanomappdx}}

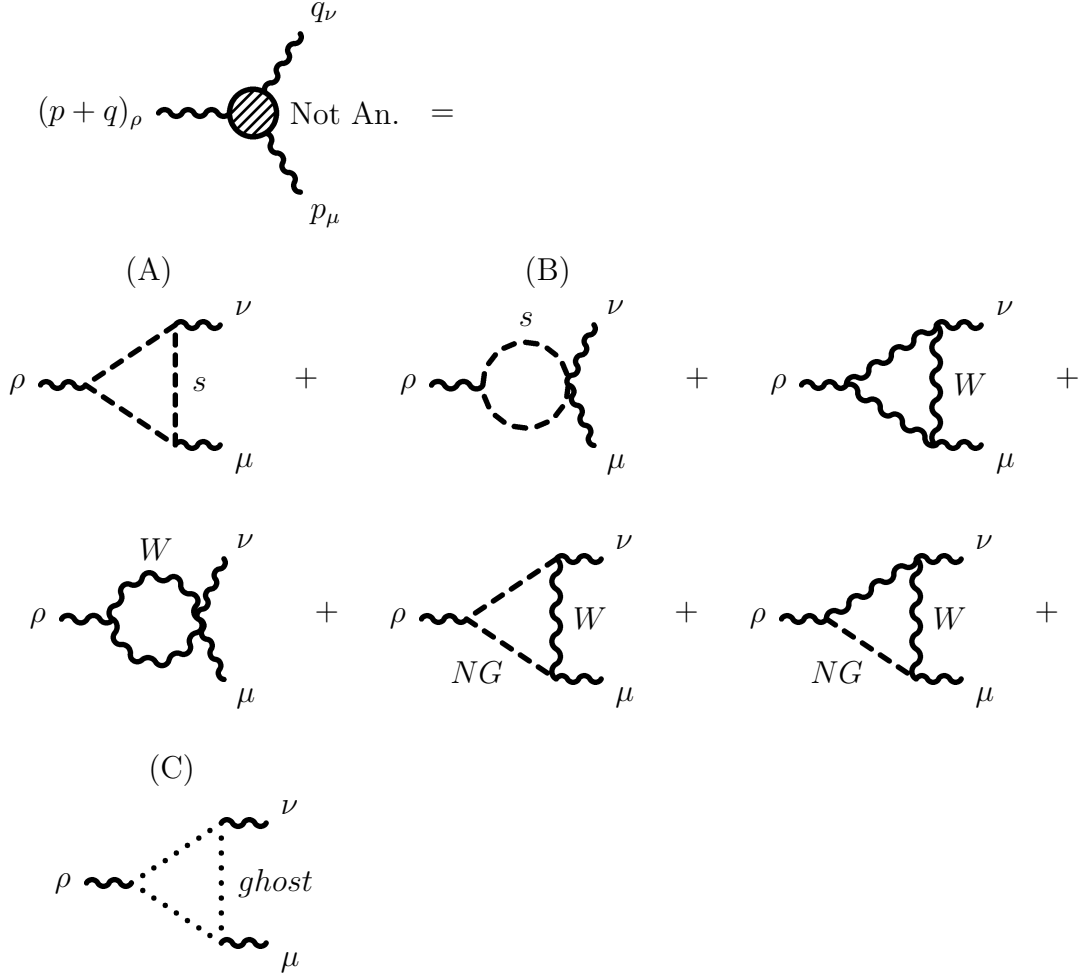
\begin{figure}[tb]
      \centering
        \vskip1.5cm
      \raisebox{-5.2ex}[0cm][0cm]{\unitlength=0.7mm
      \begin{fmffile}{BBBMSSM_NAnew}
      \begin{fmfgraph*}(30,30)
       \fmfpen{thick} \fmfleft{o1} \fmfright{i1,i2} \fmf{boson}{i1,v1,i2}\fmf{boson}{o1,v1}
       \fmfv{decor.shape=circle,decor.filled=shaded, decor.size=.30w}{v1}
       \fmflabel{$(p+q)_\r$}{o1}\fmflabel{$p_\m$}{i1}\fmflabel{$q_\n$}{i2}\fmflabel{~~Not An.}{v1}
      \end{fmfgraph*}
      \end{fmffile}}
       ~~~~~~~~~=~~~~~~~~~~~~~~~~~~~~~~~~~~~~~~~~~~~~~~~~~~~~~~~~~~~~~~

      \vskip 1.5cm
      (A)~~~~~~~~~~~~~~~~~~~~~~~~~~~~~~~~~~(B)~~~~~~~~~~~~~~~~~~~~~~~~~~~~~~~~~~~~~~~~~~~~
      \vskip 0.8cm
      \raisebox{-4.3ex}[0cm][0cm]{\unitlength=0.4mm
      \begin{fmffile}{ZZg_s_111}
      \begin{fmfgraph*}(60,40)
       \fmfpen{thick} \fmfleft{ii0,ii1,ii2} \fmfstraight \fmffreeze \fmftop{ii2,t1,t2,t3,oo2}
       \fmfbottom{ii0,b1,b2,b3,oo1} \fmf{phantom}{ii2,t1,t2} \fmf{phantom}{ii0,b1,b2} \fmf{phantom}{t1,v1,b1}
       \fmf{phantom}{t2,b2} \fmf{phantom}{t3,b3}
       \fmffreeze
       \fmf{photon}{ii1,v1} \fmf{dashes}{t3,v1} \fmf{dashes,label=$s$}{b3,t3} \fmf{dashes}{v1,b3} \fmf{photon}{b3,oo1}
       \fmf{photon}{t3,oo2}
       \fmflabel{$\r$}{ii1} \fmflabel{$\m$}{oo1} \fmflabel{$\n$}{oo2}
      \end{fmfgraph*}
      \end{fmffile}}
      ~~~~~+~~~~~
      ~~
      \raisebox{-4.3ex}[0cm][0cm]{\unitlength=0.4mm
       \begin{fmffile}{ZZg_2_17}
       \begin{fmfgraph*}(60,40)
        \fmfpen{thick} \fmfleft{ii1} \fmfright{oo1,oo2} \fmf{photon}{ii1,v1} \fmf{phantom,left,tension=0.3}{v1,v2,v1}
        \fmf{photon}{oo1,v2,oo2}\fmffreeze
        \fmf{dashes,left,tension=0.3,label=$s$}{v1,v2} \fmf{dashes,left,tension=0.3}{v2,v1}
        \fmflabel{$\r$}{ii1} \fmflabel{$\m$}{oo1} \fmflabel{$\n$}{oo2}
       \end{fmfgraph*}
       \end{fmffile}}
       ~~~~~+~~~~~~
       \raisebox{-4.3ex}[0cm][0cm]{\unitlength=0.4mm
       \begin{fmffile}{ZZg_W_111}
       \begin{fmfgraph*}(60,40)
        \fmfpen{thick} \fmfleft{ii0,ii1,ii2} \fmfstraight \fmffreeze \fmftop{ii2,t1,t2,t3,oo2}
        \fmfbottom{ii0,b1,b2,b3,oo1} \fmf{phantom}{ii2,t1,t2} \fmf{phantom}{ii0,b1,b2} \fmf{phantom}{t1,v1,b1}
        \fmf{phantom}{t2,b2} \fmf{phantom}{t3,b3}
        \fmffreeze
        \fmf{photon}{ii1,v1} \fmf{photon}{t3,v1} \fmf{photon,label=$W$}{b3,t3} \fmf{photon}{v1,b3} \fmf{photon}{b3,oo1}
        \fmf{photon}{t3,oo2}
        \fmflabel{$\r$}{ii1} \fmflabel{$\m$}{oo1} \fmflabel{$\n$}{oo2}
       \end{fmfgraph*}
       \end{fmffile}}
       ~~~~~+~~~~~
       \vskip 2.5cm
       \raisebox{-4.3ex}[0cm][0cm]{\unitlength=0.4mm
       \begin{fmffile}{ZZg_2W_17}
       \begin{fmfgraph*}(60,40)
        \fmfpen{thick} \fmfleft{ii1} \fmfright{oo1,oo2} \fmf{photon}{ii1,v1} \fmf{phantom,left,tension=0.3}{v1,v2,v1}
        \fmf{photon}{oo1,v2,oo2}\fmffreeze
        \fmf{photon,left,tension=0.3,label=$W$}{v1,v2} \fmf{photon,left,tension=0.3}{v2,v1}
        \fmflabel{$\r$}{ii1} \fmflabel{$\m$}{oo1} \fmflabel{$\n$}{oo2}
       \end{fmfgraph*}
       \end{fmffile}}
       ~~~~~+~~~~~
       \raisebox{-4.3ex}[0cm][0cm]{\unitlength=0.4mm
       \begin{fmffile}{ZZg_WGG_121}
       \begin{fmfgraph*}(60,40)
        \fmfpen{thick} \fmfleft{ii0,ii1,ii2} \fmfstraight \fmffreeze \fmftop{ii2,t1,t2,t3,oo2}
        \fmfbottom{ii0,b1,b2,b3,oo1} \fmf{phantom}{ii2,t1,t2} \fmf{phantom}{ii0,b1,b2} \fmf{phantom}{t1,v1,b1}
        \fmf{phantom}{t2,b2} \fmf{phantom}{t3,b3}
        \fmffreeze
        \fmf{photon}{ii1,v1} \fmf{dashes}{t3,v1} \fmf{photon,label=$W$}{b3,t3} \fmf{dashes,label=$NG$}{v1,b3}
        \fmf{photon}{b3,oo1} \fmf{photon}{t3,oo2}
        \fmflabel{$\r$}{ii1} \fmflabel{$\m$}{oo1} \fmflabel{$\n$}{oo2}
       \end{fmfgraph*}
       \end{fmffile}}
       ~~~~~+~~~~~
       \raisebox{-4.3ex}[0cm][0cm]{\unitlength=0.4mm
       \begin{fmffile}{ZZg_WWG_121}
       \begin{fmfgraph*}(60,40)
        \fmfpen{thick} \fmfleft{ii0,ii1,ii2} \fmfstraight \fmffreeze \fmftop{ii2,t1,t2,t3,oo2}
        \fmfbottom{ii0,b1,b2,b3,oo1} \fmf{phantom}{ii2,t1,t2} \fmf{phantom}{ii0,b1,b2} \fmf{phantom}{t1,v1,b1}
        \fmf{phantom}{t2,b2} \fmf{phantom}{t3,b3}
        \fmffreeze
        \fmf{photon}{ii1,v1} \fmf{photon}{t3,v1} \fmf{photon,label=$W$}{b3,t3} \fmf{dashes,label=$NG$}{v1,b3}
        \fmf{photon}{b3,oo1} \fmf{photon}{t3,oo2}
        \fmflabel{$\r$}{ii1} \fmflabel{$\m$}{oo1} \fmflabel{$\n$}{oo2}
       \end{fmfgraph*}
       \end{fmffile}}
       ~~~~~+~~~~~
      \vskip 1.5cm
      (C)~~~~~~~~~~~~~~~~~~~~~~~~~~~~~~~~~~~~~~~~~~~~~~~~~~~~~~~~~~~~~~~~~~~~~~~~~~~~~~
      \vskip 0.8cm
       \raisebox{-4.3ex}[0cm][0cm]{\unitlength=0.4mm
       \begin{fmffile}{ZZg_ghost_111}
       \begin{fmfgraph*}(60,40)
        \fmfpen{thick} \fmfleft{ii0,ii1,ii2} \fmfstraight \fmffreeze \fmftop{ii2,t1,t2,t3,oo2}
        \fmfbottom{ii0,b1,b2,b3,oo1} \fmf{phantom}{ii2,t1,t2} \fmf{phantom}{ii0,b1,b2} \fmf{phantom}{t1,v1,b1}
        \fmf{phantom}{t2,b2} \fmf{phantom}{t3,b3}
        \fmffreeze
        \fmf{photon}{ii1,v1} \fmf{dots}{t3,v1} \fmf{dots,label=$ghost$}{b3,t3} \fmf{dots}{v1,b3} \fmf{photon}{b3,oo1}
        \fmf{photon}{t3,oo2}
        \fmflabel{$\r$}{ii1} \fmflabel{$\m$}{oo1} \fmflabel{$\n$}{oo2}
       \end{fmfgraph*}
       \end{fmffile}}~~~~~~~~~~~~~~~~~~~~~~~~~~~~~~~~~~~~~~~~~~~~~~~~~~~~~~~~~~~~~~~~~~~~~~~~~~~~~~
       \vskip 1cm
      \caption{Non Anomalous diagrams for trilinear neutral gauge boson amplitudes.}\label{OtherDiagrams}
     \end{figure}
In this section we show that the non anomalous diagrams in Fig. \ref{diagzp} vanish. The
diagrams we consider, reported in Fig. \ref{OtherDiagrams}, have no specific assignment for the external legs, to keep
the discussion as general as possible.
All the factors which are not relevant for our aim are omitted and all the possible leg exchanges are understood.
Finally, we use dimensional regularization and the $R_\xi$ gauge with $\xi=1$, in such a way that each diagram vanishes separately.

A) The Scalar triangle loop is given by
       \bea
        D^A_{\m \n \r} (p,q) &=&  \int \frac{d^{2 \w} l}{(2 \pi)^{2 \w}}
        \frac{(2l+p-q)_\r (2l-q)_\n (2l+p)_\m}{\[(l-q)^2-m^2\]\[l^2-m^2\]\[(l+p)^2-m^2\]}
        + \, (p \leftrightarrow q, \m \leftrightarrow \n) \nn\\
        &=& \int \frac{d^{2 \w} l}{(2 \pi)^{2 \w}}
        \frac{(2l+p-q)_\r (2l-q)_\n (2l+p)_\m}{\[(l-q)^2-m^2\]\[l^2-m^2\]\[(l+p)^2-m^2\]} \nn\\
        &&+\int \frac{d^{2 \w} l}{(2 \pi)^{2 \w}}
        \frac{(2l+q-p)_\r (2l-p)_\m (2l+q)_\n}{\[(l-p)^2-m^2\]\[l^2-m^2\]\[(l+q)^2-m^2\]}
       \eea
 Performing the change of variable $l_\m \to -l_\m$ in the second integral, one gets
       \bea
        D^A_{\m \n \r} (p,q)
        &=& \int \frac{d^{2 \w} l}{(2 \pi)^{2 \w}}
        \frac{(2l+p-q)_\r (2l-q)_\n (2l+p)_\m}{\[(l-q)^2-m^2\]\[l^2-m^2\]\[(l+p)^2-m^2\]} \nn\\
        &&-\int \frac{d^{2 \w} l}{(2 \pi)^{2 \w}}
        \frac{(2l+p-q)_\r (2l+p)_\m (2l-q)_\n}{\[(l-q)^2-m^2\]\[l^2-m^2\]\[(l+p)^2-m^2\]} = 0
       \label{diaga}\eea

 B) The ``Scalar bubble loop'' is given by
       \bea
        D^B_{\m \n \r} (p,q)
        &=& -2 \int \frac{d^{2 \w} l}{(2 \pi)^{2 \w}}
        \frac{(2l+p+q)_\r \emn}{\[l^2-m^2\]\[(l+p+q)^2-m^2\]} \nn\\
        &=& -2 \int \frac{d^{2 \w} l}{(2 \pi)^{2 \w}}
        \frac{(l+p+q)_\r \emn}{\[l^2-m^2\]\[(l+p+q)^2-m^2\]}\nn\\
        &&-2 \int \frac{d^{2 \w} l}{(2 \pi)^{2 \w}}
        \frac{(l)_\r \emn}{\[l^2-m^2\]\[(l+p+q)^2-m^2\]}
       \eea
 Performing the change of variable $l \to -l-p-q$ in the second integral one gets
       \bea
        D^B_{\m \n \r} (p,q)
        &=& -2 \int \frac{d^{2 \w} l}{(2 \pi)^{2 \w}}
        \frac{(l+p+q)_\r \emn}{\[l^2-m^2\]\[(l+p+q)^2-m^2\]}\nn\\
        &&+2 \int \frac{d^{2 \w} l}{(2 \pi)^{2 \w}}
        \frac{(l+p+q)_\r \emn}{\[(l+p+q)^2-m^2\]\[l^2-m^2\]}= 0
       \label{diagb}\eea

C) Since the ghost interact with neutral vectors only through the third component of $SU(2)$,
the Ghost triangle loop is proportional to
       \be
        \e_{3bc} \e_{3cd} \e_{3db} = -\d_{bd} \e_{3db} = 0
        \label{diagc}
       \ee
The other diagrams in Fig. \ref{OtherDiagrams} can also be shown to vanish after manipulations similar to the ones
used in (\ref{diaga}), (\ref{diagb}), (\ref{diagc}).

\section{Decay rates. General case}

In this Section we compute the amplitudes for the decays $Z'
\to Z_0 \, \g$ and $Z' \to Z_0 \, Z_0$ in the general case $Q_{H_u}\neq 0$, still neglecting the effects coming from the kinetic mixing.  We work in the limit
\be g_a
  v_{u,d}<<\m,M_0,M_1,M_2,M_S, M_{V^{(0)}}  \label{nosusylimit}
\ee
in which $m_{SC}\approx M_{V^{(0)}},\, m_{SB}\approx 0$ (see (\ref{msc}), (\ref{cdelta}),
(\ref{xdelta})). Hence,  (\ref{massmatrix}) takes the same form as in the symmetric phase in which
neutralinos and charginos do not contribute to the anomaly (see Section \ref{AnomalyCancellationSymmPhase}).
In the limit (\ref{nosusylimit}) an extension of the standard model by an extra $U(1)$ and our SUSY model
give the same results for what the decays of interest are concerned.

We define the Dirac fermions $ \Psi_f = \( \begin{array}{c} f_L\\
                                                            f_R \end{array} \)$
where $f_{L(R)}$ are all the left(right) Weyl fermions in the model. The SM
fermion interaction terms with the neutral gauge bosons are
\bea
 \L^{int}_{Z'} &=& J^\m_{Z'} Z'_\m \nn = - {1\over2} \, g_{Z'} \, \bar \Psi_f \, \g^\m \( v_f^{Z'} - a_f^{Z'} \g_5 \) \Psi_f Z'_\m\\
 \L^{int}_{Z_0} &=& J^\m_{Z_0} Z_{0 \m} =- {1\over2} \, g_{Z_0} \, \bar \Psi_f \, \g^\m \( v_f^{Z_0} - a_f^{Z_0} \g_5 \) \Psi_f Z_{0 \m} \nn\\
 \L^{int}_{\g} &=& J^\m_{\g} A_\m =- e \, q_f \bar \Psi_f \, \g^\m \Psi_f A_\m  \label{neutralcurrents}
\eea
where
\bea
 &&v_f^{Z'} = Q^{Z'}_{f_L}+Q^{Z'}_{f_R}  \quad\qquad
   a_f^{Z'} = Q^{Z'}_{f_L}-Q^{Z'}_{f_R}  \nn\\
 &&v_f^{Z_0}= Q^{Z_0}_{f_L}+Q^{Z_0}_{f_R}  \quad\qquad
   a_f^{Z_0}= Q^{Z_0}_{f_L}-Q^{Z_0}_{f_R} \nn\\
 &&\ q_f      = Q_{f_L}=Q_{f_R}
\eea
The left and right charges are defined in the following way
\bea
g_{Z'} Q^{Z'}_{f_L} &=& g_2 T_3 O_{02} + g_1 Y_{f_L} O_{01} + g_0 Q_{f_L} \\
g_{Z'} Q^{Z'}_{f_R} &=& g_1 Y_{f_R} O_{01} + g_0 Q_{f_R} \\
g_{Z_0} Q^{Z_0}_{f_L} &=& g_2 T_3 O_{12} + g_1 Y_{f_L} O_{11} + g_0 Q_{f_L} O_{10}\\
g_{Z_0} Q^{Z_0}_{f_R} &=& g_1 Y_{f_R} O_{11} + g_0 Q_{f_R} O_{10} \\
e Q_{f_L} &=& g_2 T_3 O_{22} + g_1 Y_{f_L} O_{21} = g_1 Y_{f_R} O_{21} = e Q_{f_R}
\eea
where $O_{ij}$ is given in (\ref{Oij}) and $T_3$ is the eigenvalue of $T^{(2)}_3$.

\subsection{$Z' \to Z_0 \ \g$}
The amplitude is given by the sum of the fermionic triangle $\D_{\r \m \n}^{Z' Z_0 \g}$ plus the proper GCS vertex    %
    \bea
     A_{\r \m \n}^{Z' Z_0 \g}&=& \D_{\r \m \n}^{Z' Z_0 \g} + (GCS)_{\r \m \n}^{Z' Z_0 \g} \nn\\
             \D_{\r \m \n}^{Z' Z_0 \g}&=& -{1\over4} g_{Z'} g_{Z_0} e \sum_f \( v_f^{Z'} a_f^{Z_0} q_f \G^{VAV}_{\r \m \n} +
                                                                               a_f^{Z'} v_f^{Z_0} q_f \G^{AVV}_{\r \m \n} \)
    \eea
The resulting amplitude can be written as
    \bea
     A_{\r \m \n}^{Z' Z_0 \g}&=&
        -{1\over4\pi^2} g_{Z'} g_{Z_0} e  \Big(\tilde A_1 \e[p,\m,\n,\r] +
         \tilde A_2\e[q,\m,\n,\r]+ A_3 \e[p,q,\m,\r]{p}^{\n}\nn\\
        &&+  A_4 \e[p,q,\m,\r]{q}^{\n} +  A_5 \e[p,q,\n,\r]p^\m +
         A_6\e[p,q,\n,\r]q^\m \Big)
      \eea
with
\be
 A_i=\sum_f \(v_f^{Z'} a_f^{Z_0}+a_f^{Z'} v_f^{Z_0}\) q_f I_i  \qquad \text{for } i=3, \dots , 6
\ee
 and the integrals $I_i$ given in (\ref{I's}). $\tilde A_1$ and $\tilde A_2$ are the new coefficients with the GCS absorbed similarly to (\ref{I1tilde}).

The Ward identities (\ref{GoldWI}) for the
amplitude now read
     \bea
     (p+q)^\r A_{\r \m \n}^{Z' Z_0 \g}+i M_{Z'} \[(GS)^{Z_0 \g}_{\m \n}+(NG)^{Z_0 \g}_{\m \n}\]&=&0\label{WI-ZZg-general1}\\
      p^\m A_{\r \m \n}^{Z' Z_0 \g}+i M_{Z_0} \[ (GS)^{Z' \g}_{\r \n}+(NG)^{Z' \g}_{\r \n}\]&=&0\label{WI-ZZg-general2}\\
      q^\n  A_{\r \m \n}^{Z' Z_0 \g}&=& 0
    \label{WI-ZZg-general3}\eea
where $M_{Z'}$ and $M_{Z_0}$ are the $Z'$ and $Z_0$ masses
respectively.
In both (\ref{WI-ZZg-general1}) and (\ref{WI-ZZg-general2}) we have a $(GS)$ and a
$(NG)$ contribution due to the two Goldstone bosons which are a
linear combination of the axion and $G^0$.
We use (\ref{WI-ZZg-general2}) and (\ref{WI-ZZg-general3}) to fix $\tilde A_1$ and $\tilde A_2$ while
(\ref{WI-ZZg-general1}) is automatically satisfied. Contracting with $p^\mu$ we get
    \bea
     p^\m A_{\r \m \n}^{Z' Z_0 \g}&=&-\Bigg\{8 \[ 4 g_0 g_1^2 \ R_{101}^{Z' Z_0 \g} \ b_2^{(1)} b_3 +
                                                 2 g_0 g_2^2 \ R_{202}^{Z' Z_0 \g} \ b_2^{(2)} b_3 +
                                                 2 g_0^2 g_1 \ R_{001}^{Z' Z_0 \g} \ b_2^{(4)} b_3\] +\nn\\
                                   &&~~~~~~+ {1\over4\pi^2} g_{Z'} g_{Z_0} e \sum_f v_f^{Z'} a_f^{Z_0} q_f \ m_f^2 I_0                                    \Bigg\} \ \e[q,p,\n,\r]
    \eea
    where $I_0$ is the integral given in (\ref{I_0integral}).
    The solution for $\tilde A_1$ and $\tilde A_2$ is
    \bea
       &&\tilde A_1      =       \(q^2 A_4 + p \cdot q A_3 \) \\
       &&\tilde A_2      =       \(p^2 A_5 + p \cdot q A_6\) +(GS)^{Z' \g} +(NG)^{Z' \g}
    \eea
 with
\bea
 (NG)^{Z' \g}&=&\sum_f v_f^{Z'} a_f^{Z_0} q_f \ m_f^2 I_0 \\
 (GS)^{Z' \g}&=&\frac{32 \pi^2}{g_{Z'} g_{Z_0} e} \[ 4 g_0 g_1^2 \ R_{101}^{Z' Z_0 \g} \ b_2^{(1)} b_3 +
                                                 2 g_0 g_2^2 \ R_{202}^{Z' Z_0 \g} \ b_2^{(2)} b_3 +
                                                 2 g_0^2 g_1 \ R_{001}^{Z' Z_0 \g} \ b_2^{(4)} b_3\]\nn\\
\eea
The rotation factors are
\bea
 R_{101}^{Z' Z_0 \g}       &=& O_{01} O_{10} O_{21} \nn\\
 R_{202}^{Z' Z_0 \g} &=& O_{02} O_{10} O_{22} \nn\\
 R_{001}^{Z' Z_0 \g}      &=& O_{10} O_{21}
\eea
with $O_{ij}$ given by (\ref{Oij}). Substituting $\tilde A_1, ~ \tilde A_2$ into the amplitude (\ref{ampZ'Zg}) and performing all the contractions
we finally obtain
    \bea
     &&|A_{\text{TOT}}|^2_{Z' Z_0 \g} = g_{Z'}^2 g_{Z_0}^2 e^2 \frac{
     \left(M_{Z'}^2-M_{Z_0}^2\right)^2 \left(M_{Z'}^2+M_{Z_0}^2\right)}{96 M_{Z_0}^2 M_{Z'}^2 \pi ^4} \times
    ~~~~~~~~~~~~~~~~~~~~\nn\\
    &&~~~~~~~~~~~~~~
    \[\sum_f q_f \(v_f^{Z'} a_f^{Z_0}+a_f^{Z'} v_f^{Z_0}\) (I_3+I_5) M_{Z_0}^2
     +(GS)^{Z' \g} +(NG)^{Z' \g}  \]^2
   \eea

\subsection{$Z' \to Z_0 \ Z_0$}
The contribution to the fermionic triangle is

\bea
 &&\D_{\r \m \n}^{Z' Z_0 Z_0}=-{1\over8} g_{Z'} g_{Z_0}^2 \Bigg[\sum_f \(
     v_f^{Z'} a_f^{Z_0} v_f^{Z_0} \, \G_{\r \m \n}^{VAV} +
     v_f^{Z'} v_f^{Z_0} a_f^{Z_0} \, \G_{\r \m \n}^{VVA} +
     a_f^{Z'} v_f^{Z_0} v_f^{Z_0} \, \G_{\r \m \n}^{AVV} \right.  + \nn\\
    &&~~~~~~~~~~~~~~~~~~~~~~~~~~~~~~~~~~~ +  \left.
     a_f^{Z'} a_f^{Z_0} a_f^{Z_0} \, \G_{\r \m \n}^{AAA} \) \Bigg]
\eea
where the $\G_{\r \m \n}$'s are given by (\ref{AVVtrian}), (\ref{AAAtrian}), (\ref{VAVtrian}), (\ref{VVAtrian}).
We write the total amplitude (the sum of the triangles plus GCS terms) as
\bea
     A_{\r \m \n}^{Z' Z_0 Z_0}&=&
        -{1\over8\pi^2} g_{Z'} g_{Z_0}^2 \Big[\tilde A_1 \e[p,\m,\n,\r] +
         \tilde A_2\e[q,\m,\n,\r]+
          A_3 \e[p,q,\m,\r]{p}_{\n} \nn\\
        &&+  A_4 \e[p,q,\m,\r]{q}_{\n} +  A_5 \e[p,q,\n,\r]p_\m +
         A_6\e[p,q,\n,\r]q_\m \Big]
\eea
with
\be
 A_i=\sum_f t_f^{Z' Z_0 Z_0} I_i  \qquad \text{for } i=3, \dots , 6
\ee
where
\be
 t_f^{Z' Z_0 Z_0}  =   \( a_f^{Z'} v_f^{Z_0} v_f^{Z_0}+ 2 v_f^{Z'} a_f^{Z_0} v_f^{Z_0}  + a_f^{Z'} a_f^{Z_0} a_f^{Z_0}\)
\ee
and the integrals $I_i$ are given in (\ref{I's}). The Ward identities now read
     \bea
     (p+q)^\r A_{\r \m \n}^{Z' Z_0 Z_0} +i M_{Z'} \[ (GS)^{Z_0 Z_0}_{\m \n}+(NG)^{Z_0 Z_0}_{\m \n} \]&=&0\label{WI-ZZZ-general1}\\
      p^\m A_{\r \m \n}^{Z' Z_0 Z_0} +i M_{Z_0} \[ (GS)^{Z' Z_0}_{\r \n}+(NG)^{Z' Z_0}_{\r \n} \]&=&0\label{WI-ZZZ-general2}\\
      q^\n A_{\r \m \n}^{Z' Z_0 Z_0} +i M_{Z_0} \[ (GS)^{Z_0 Z'}_{\m \r}+(NG)^{Z_0 Z'}_{\m \r} \]&=&0
   \label{WI-ZZZ-general3} \eea
where $M_{Z'}$ and $M_{Z_0}$ are the $Z'$ and $Z_0$ masses respectively.
In (\ref{WI-ZZZ-general1})-(\ref{WI-ZZZ-general3}) the $(GS)$ and $(NG)$ terms are present for the same reason
as in the preceding Subsection.
We use (\ref{WI-ZZZ-general2}) and (\ref{WI-ZZZ-general3}) to fix $\tilde A_1$ and $\tilde A_2$ while
(\ref{WI-ZZZ-general1}) is automatically satisfied.

Contracting with $p^\mu$ and $q^\nu$ we get
    \bea
     p^\m A_{\r \m \n}^{Z' Z_0 Z_0}&=&-\Bigg\{8 \[4 g_0^3 \ R_{000}^{Z' Z_0 Z_0} \ b_2^{(0)} b_3 + 4 g_0 g_1^2 \ R_{101}^{Z' Z_0 Z_0} \ b_2^{(1)} b_3 +\right.\nn\\
                                                &&~~~~~~~~\left.+ 2 g_0 g_2^2 \ R_{202}^{Z' Z_0 Z_0} \ b_2^{(2)} b_3 +
                                                 2 g_0^2 g_1 \ R_{001}^{Z' Z_0 Z_0} \ b_2^{(4)} b_3\] +\nn\\
     &&~~~~~~+ {1\over8\pi^2} g_{Z'} g_{Z_0}^2 \sum_f \(v_f^{Z'} a_f^{Z_0} v_f^{Z_0} +{1\over3} a_f^{Z'} a_f^{Z_0} a_f^{Z_0} \)\ m_f^2 I_0                                    \Bigg\} \ \e[q,p,\n,\r] \qquad \quad\\
     q^\n A_{\r \m \n}^{Z' Z_0 Z_0}&=&-\Bigg\{8 \[4 g_0^3 \ R_{000}^{Z' Z_0 Z_0} \ b_2^{(0)} b_3 + 4 g_0 g_1^2 \ R_{101}^{Z' Z_0 Z_0} \ b_2^{(1)} b_3 +\right.\nn\\
                                                &&~~~~~~~~\left.+ 2 g_0 g_2^2 \ R_{202}^{Z' Z_0 Z_0} \ b_2^{(2)} b_3 +
                                                 2 g_0^2 g_1 \ R_{001}^{Z' Z_0 Z_0} \ b_2^{(4)} b_3\] +\nn\\
     &&~~~~~~+ {1\over8\pi^2} g_{Z'} g_{Z_0}^2 \sum_f \(v_f^{Z'} a_f^{Z_0} v_f^{Z_0} +{1\over3} a_f^{Z'} a_f^{Z_0} a_f^{Z_0} \)\ m_f^2 I_0                                    \Bigg\} \ \e[q,p,\r,\m]
    \eea
where $I_0$ is the integral given in (\ref{I_0integral}).
    The solution for $\tilde A_1$ and $\tilde A_2$ is
    \bea
       &&\tilde A_1      =        \(q^2 A_4 + p \cdot q A_3 \) -\[ (GS)^{Z' Z_0}+(NG)^{Z' Z_0} \]\\
       &&\tilde A_2       =       \(p^2 A_5 + p \cdot q A_6\) +(GS)^{Z' Z_0}+(NG)^{Z' Z_0}
    \eea
 with
\bea
 (NG)^{Z' Z_0}&=&\sum_f \(v_f^{Z'} a_f^{Z_0} v_f^{Z_0} +{1\over3} a_f^{Z'} a_f^{Z_0} a_f^{Z_0} \)\ m_f^2 I_0 \\
 (GS)^{Z' Z_0}&=&\frac{64 \pi^2}{g_{Z'} g_{Z_0}^2} \Bigg[ 4 g_0^3 \ R_{000}^{Z' Z_0 Z_0}     \ b_2^{(0)} b_3 +
                                                       4 g_0 g_1^2 \ R_{101}^{Z' Z_0 Z_0}       \ b_2^{(1)} b_3
                                                       +\nn\\
                                     && \qquad \quad  +2 g_0 g_2^2 \ R_{202}^{Z' Z_0 Z_0} \ b_2^{(2)} b_3 +
                                                       2 g_0^2 g_1 \ R_{001}^{Z' Z_0 Z_0}      \ b_2^{(4)} b_3
                                                       \Bigg]
\eea
The rotation factors are
\bea
 R_{000}^{Z' Z_0 Z_0}     &=& O_{10} O_{10} \nn\\
 R_{101}^{Z' Z_0 Z_0}       &=& O_{01} O_{10} O_{11} \nn\\
 R_{202}^{Z' Z_0 Z_0} &=& O_{02} O_{10} O_{12} \nn\\
 R_{001}^{Z' Z_0 Z_0}      &=& O_{10} O_{11} + O_{01} O_{10} O_{10}
\eea
with $O_{ij}$ given by (\ref{Oij}). Substituting back into the
amplitude and performing all the contractions we finally obtain
\bea
 |A_{\text{TOT}}|^2_{Z' Z_0 Z_0} &=&
 g_{Z'}^2 g_{Z_0}^4 \frac{ \left(M_{Z'}^2-4 M_{Z_0}^2\right)^2}{192 M_{Z_0}^2\pi^4}
 \times \\
 && \[ \sum_f t_f^{Z' Z_0 Z_0} (I_3+I_5) M_{Z_0}^2 + (GS)^{Z' Z_0}+(NG)^{Z' Z_0}\]^2  \nn\\
\eea

\end{document}